\documentclass[fleqn,usenatbib]{mnras}

\usepackage{newtxtext,newtxmath}
\usepackage[T1]{fontenc}

\DeclareRobustCommand{\VAN}[3]{#2}
\let\VANthebibliography\thebibliography
\def\thebibliography{\DeclareRobustCommand{\VAN}[3]{##3}\VANthebibliography}

\usepackage{graphicx}
\usepackage{natbib}
\usepackage{enumerate}
\usepackage[export]{adjustbox}
\usepackage{amsmath, amsfonts}
\usepackage{gensymb}
\usepackage{mathtools}
\usepackage{natbib}
\usepackage{color}
\usepackage{hyperref}
\usepackage{ulem}
\usepackage{soul}
\usepackage{url}
\usepackage{float}
\usepackage{xspace}
\usepackage{comment}
\usepackage{longtable}
\usepackage{array}
\usepackage{xspace}
\usepackage{cellspace}
\newcommand{\Targetname}{HD~63754\xspace}
\newcommand{\HIPname}{HIP~38216\xspace}
\newcommand{\Hipparcos}{{\sl Hipparcos}}
\newcommand{\Gaia}{{\sl Gaia}\xspace}

\newcommand{\Msun}{\mbox{$M_{\sun}$}}

\newcommand{\Lsun}{\mbox{$L_{\sun}$}}

\newcommand{\Mjup}{\mbox{$M_{\rm Jup}$}}

\newcommand{\orvara}{{\tt orvara }}
\newcommand{\VIP}{{\texttt{VIP}}}

\title[]{The Keck-HGCA Pilot Survey II: Direct Imaging Discovery of \Targetname~B, a $\sim$20 au Massive Companion Near the Hydrogen Burning Limit}

\author[Y.~Li et al.]{Yiting Li$^{1,8}$\thanks{E-mail: lyiting@umich.edu},
Timothy D.~Brandt$^{1}$,
Kyle Franson$^{2}$,
Qier An$^{1}$,
Taylor Tobin $^{8}$,
Thayne Currie$^{7}$,
Minghan Chen$^{1}$,
\newauthor
Lanxuan Wang$^{1}$,
Trent J.~Dupuy$^{3}$, 
Rachel Bowens-Rubin$^{4}$,
Ma\"issa Salama$^{4}$,
Briley L.~Lewis$^{5}$,
\newauthor
Aidan Gibbs$^{5}$,
Brendan P.~Bowler$^{2}$
Rebecca Jensen-Clem$^{4}$,
Jacqueline Faherty$^{6}$, 
Michael P.~Fitzgerald$^{5}$,
\newauthor
Benjamin A.~Mazin$^{1}$
\\
$^{1}$Department of Physics, University of California, Santa Barbara, Santa Barbara, CA 93106, USA\\
$^{2}$Department of Astronomy, The University of Texas at Austin, Austin, TX 78712, USA \\
$^{3}$Institute for Astronomy, University of Edinburgh, Royal Observatory, Blackford Hill, Edinburgh, EH9 3HJ, UK \\
$^{4}$Astronomy \& Astrophysics Department, University of California, Santa Cruz, CA 95064, USA \\
$^{5}$Department of Physics and Astronomy, University of California, Los Angeles, 475 Portola Plaza, Los Angeles, CA 90025 \\
$^{6}$American Museum of Natural History, New York, NY, USA \\
$^{7}$The University of Texas at San Antonio, San Antonio, TX 78249\\
$^{8}$Department of Astronomy, University of Michigan, Ann Arbor, MI 48109
}

\date{Accepted XXX. Received YYY; in original form ZZZ}

\pubyear{2023}

\begin{document}
\label{firstpage}
\pagerange{\pageref{firstpage}--\pageref{lastpage}}
\maketitle

\begin{abstract}

We present the joint astrometric and direct imaging discovery, mass measurement, and orbital analysis of \Targetname~B (\HIPname~B), a companion near the stellar-substellar boundary orbiting $\sim$20 AU from its Sun-like host. \Targetname was observed in our ongoing high-contrast imaging survey targeting stars with significant proper-motion accelerations between \Hipparcos and \Gaia consistent with wide-separation substellar companions. We utilized archival HIRES and HARPS radial velocity (RV) data, together with the host star's astrometric acceleration extracted from the \Hipparcos--\Gaia Catalog of Accelerations (HGCA), to predict the location of the candidate companion around \Targetname~A. We subsequently imaged \Targetname~B at its predicted location using the Near Infrared Camera 2 (NIRC2) in the $L'$ band at the W.~M.~Keck Observatory. We then jointly modeled the orbit of \Targetname~B with RVs, \Hipparcos--\Gaia accelerations,  and our new relative astrometry, measuring a dynamical mass of  ${81.9}_{-5.8}^{+6.4} \Mjup$,  an eccentricity of ${0.260}_{-0.059}^{+0.065}$, and a nearly face-on inclination of $174.\!\!^\circ81_{-0.50}^{+0.48}$. For \Targetname~B, we obtain an $L'$ band absolute magnitude of $L' = 11.39\pm0.06$ mag, from which we infer a bolometric luminosity of $\mathrm{log(L_{\rm bol}/\Lsun)= -4.55 \pm0.08}$ dex using a comparison sample of L and T dwarfs with measured luminosities. Although uncertainties linger in age and dynamical mass estimates, our analysis points toward \Targetname~B's identity as a brown dwarf on the L/T transition rather than a low-mass star, indicated by its inferred bolometric luminosity and model-estimated effective temperature. Future RV, spectroscopic, and astrometric data such as those from JWST and Gaia DR4 will clarify \Targetname~B's mass, and enable spectral typing and atmospheric characterization. 

\\ \\
\end{abstract}

\section{Introduction}

Brown dwarfs are distinct from stars in that they do not have enough mass to sustain hydrogen fusion in their cores to counteract gravitational collapse. Instead, as they evolve, they are supported first by thermal pressure from a combination of deuterium burning and gravitational contraction, and ultimately by electron degeneracy pressure \citep{Henriksen_1986,Chabrier_2005,McKee_2007,Andre_2009,Dobbs_2013,Schneider_2015,Forbes_2019, Nony_2023}. As a result of heat loss, they continuously cool across the L, T, Y spectrum and emit their light in the thermal infrared \citep{Lunine_1986}. Brown dwarfs manifest temperatures lower than the coolest stars, occupying a range of surface temperatures, spanning 3000 K for the hottest L dwarfs \citep{Kirkpatrick_2000,Filippazzo_2015} to a frosty 250 K for the coolest Y dwarfs \citep{Katharina_2002,Beichman_2013}. However, recent discoveries, such as an 8000 K irradiated-Jupiter-analogue brown dwarf orbiting a hot white dwarf \citep{Hallakoun_2023}, continue to challenge our understanding of the rich tapestry of brown dwarf diversity. 

The mechanisms governing brown dwarf formation are still not fully understood, with several proposed channels \citep{Whitworth_2018}, including turbulent fragmentation \citep{Padoan_2002,Padoan_2004}, fragmentation within filaments and discs \citep{Peretto_2013,Balfour_2015}, dynamical ejection of protostellar embryos \citep{Reipurth_2001}, and photoerosion \citep{Hester_1996}. Observations are crucially needed to assess their relative importance. Substellar evolutionary models incorporating synthetic spectra, nongray atmosphere models, dust formation, and opacity have been developed to explain brown dwarfs' behavior in optical and infrared color-magnitude diagrams \citep{Chabrier_2000}. These evolutionary models increasingly emphasize the impact of dynamic processes on internal structure and methane absorption in the atmosphere, which affects atmospheric opacity, as effective temperatures and radii decrease in brown dwarf evolution. According to the majority of evolutionary models that distinguish between stars and brown dwarfs by their physical attributes and internal chemical processes \citep{Leggett_1998,Levine_2006,Allard_2012,Gonzales_2020}, the substellar-stellar boundary occurs near 75\,$M_{\rm Jup}$  \citep{Dieterich_2014,Pinochet_2019}. However, \citet{Chabrier_2023} recently incorporated a new Equation Of State (EOS) for dense hydrogen-helium mixtures to model the evolution of brown dwarfs and very low-mass stars, and found improved agreement with observationally determined brown dwarf masses. They report a slight increase in the hydrogen-burning minimum mass from the 75 $\Mjup$ by \citet{Dieterich_2014} to 78.5$\Mjup$. Furthermore, statistical studies on the eccentricity and obliquity of directly imaged substellar companions around cool stars have revealed a common occurrence of spin misalignments, particularly favored by stars hosting brown dwarfs over giant planets \citep{Bowler_2020, Bowler_2023}. \citep{Bowler_2020, Bowler_2023}. These findings present promising opportunities for testing distinctions between brown dwarf and giant planet formation mechanisms.

Empirical observations of the diverse substellar population anchor our physical understanding. Brown dwarfs with observationally constrained masses, ages, and luminosities independent of models offer a means to establish benchmarks for calibrating and testing evolutionary models \citep{Liu_2002,Dupuy_2009,Crepp_2014,Crepp_2014_2,Dupuy_2017,Dupuy_2019,Franson_2023}. As an example, using very precise dynamical masses in the $\varepsilon$~Indi~BC system, \citet{Chen_2022} found better agreement with evolutionary models that had slowed cooling rate near the L/T transition due to clouds and opacity effects in the atmospheres of brown dwarfs.  

The high-contrast imaging technique has emerged as a potent method for detecting exoplanets, providing spatially resolved observations of brown dwarfs. However, first-generation direct imaging searches, which adopt a blind approach by targeting every young star, have yielded low detection rates \citep{Bowler_2016,Nielsen_2019}. In recent years, attention has shifted towards targeted searches that focus on stars exhibiting stellar reflex motion, resulting in enhanced survey outcomes. Both radial velocity and astrometry techniques probe the reflex motion of stars in orthogonal directions. Radial velocity trends have been used to identify direct imaging follow-up companions \citep{Crepp_2014_2,Rickman_2019}, while astrometry trends as a target selection tool is a novel method that has been demonstrated to be effective by several surveys \citep{Currie_2023, Franson_2023}. The Hipparcos-Gaia Catalog of Accelerations (HGCA) \citep{Brandt_2021} offers absolute astrometry for about 115,000 nearby stars, including those with clear evidence of massive, unseen companions. HGCA-derived accelerations can provide dynamical masses of imaged exoplanets and low-mass brown dwarfs independently of luminosity evolution models and stellar age uncertainties \citep{Brandt_2019, Dupuy_2019}. 

In this paper, we report the discovery of a massive companion close to the hydrogen burning limit, at a separation of $0.\!\!''48$ from the Sun-like star \Targetname~A. The discovery of this companion was achieved within an ongoing survey concentrated on detecting substellar companions orbiting accelerating stars. The observations utilized infrared imaging through the NIRC2 instrument at Keck and the CHARIS integral field spectrograph at SCExAO/Subaru in Hawaii. This builds on our previous success in the survey, which yielded the discovery of a massive brown dwarf, HD~176535~B, around a K-type star \citep{Li_2023}. We present the dynamical mass determinations and orbital constraints for \Targetname~B using a combination of radial velocity, relative astrometry from direct imaging, and Hipparcos-Gaia DR3 accelerations. In Section~\ref{sec:properties}, we discuss the stellar properties of the host star \Targetname~A. Following that, we describe the prediction of the companion for our survey program in Section~\ref{sec:prediction}. We present our observation and direct imaging discovery in Section~\ref{sec:observation}. We discuss the modelings of the evolutionary properties and orbit of \Targetname~B in Section~\ref{sec:orbit_constrain} and Section~\ref{sec:discussion}, and conclude in Section~\ref{sec:conclusion}.

\section{System Properties}
\label{sec:properties}

\begin{figure}
    \centering
\includegraphics[width=0.4\textwidth]{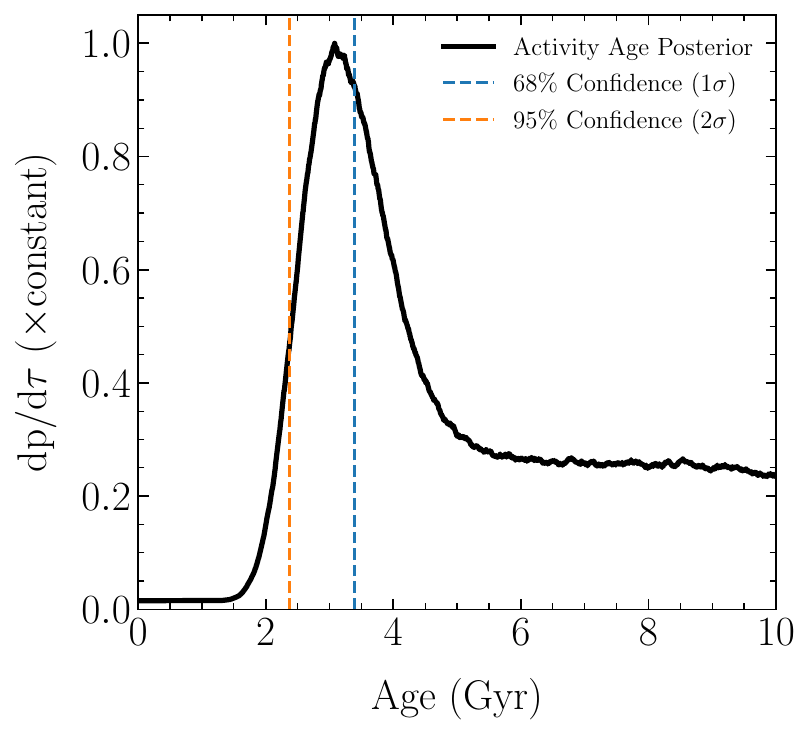}
    \caption{The normalized age probability posteriors of \Targetname~A using the Bayesian activity-age dating technique developed by \citet{Brandt_2014}. Given that the upper limit of the age is unconstrained, one-sided quantiles were calculated by integrating the area under the cumulative density function up to the 68$\%$ and 95$\%$ confidence levels, corresponding to the 1$\sigma$ and 2$\sigma$ confidence intervals, respectively. The age of \Targetname~A is estimated to be $>3.4$ Gyr at a 68$\%$ confidence level (blue dashed line), and $>2.4$ Gyr at a 95$\%$ confidence level (orange dashed line).
    \label{age_posterior}}
\end{figure}

\Targetname~A (\HIPname~A; HR~3048~A) is a bright (V$\approx$6.98 mag; \citealt{Hipparcos_1997}), main-sequence star of spectral type G0V \citep{Soubiran_2018} located at a distance of 50.17$\pm$0.05~pc \citep{GaiaEDR3_2020} in the constellation Puppis. \Targetname\ is slightly metal-rich with a metallicity of $\rm [Fe/H] = 0.20\pm0.03$ dex \citep{Brewer_2016}, a surface gravity of $\log(g) = 4.04\pm0.06$ dex \citep{Soubiran_2016}, an effective temperature of 6088$\pm$32 K \citep{Soubiran_2022}, and a luminosity of $\rm 4.88\pm0.01\,\Lsun$ based on its Gaia G magnitude \citep{GaiaCat_2022}. With above solar luminosity and effective temperature, \Targetname~A appears slightly evolved as a G-type main-sequence star. A summary of the stellar properties of \Targetname~A can be found in Table~\ref{stellar_property}.

\begin{table}
\caption{Stellar Characteristics of \Targetname~A.\label{stellar_property}}
\centering
\begin{tabular}{ccc} \hline
\textbf{Property} & \textbf{Value} & \textbf{Refs} \\ \hline
\multicolumn{3}{c}{Host Star}\\
\hline
$\varpi$ (mas)          & $19.93 \pm 0.02$      &1  \\
Distance (pc)           & $50.17 \pm 0.03$      &1  \\
SpT                     & G0V                   &2\\
Mass ($\Msun$)          & $1.41 \pm 0.15$       &13\\
Age (Gyr)               & $> 3.4$               &13\\
Radius ($R_{\odot}$)    & $1.891\pm0.035$       &8\\
$\rm T_{eff}$ (K)       & $6153 \pm 100$	    &3,4 \\
$\rm [Fe/H]$ (dex)      & $0.20 \pm 0.03$       &3,4 \\
$\rm log(g)$ (dex)      & $4.04 \pm 0.06$       &3,4\\
$\rm log(R'_{HK})$ (dex)& $-5.12\pm0.07$        &5,6,7,8 \\ 
$\rm R'_{X}$ (dex)      & $<-4.28$              &9 \\
Gaia RUWE               & 1.023                 &1 \\
Gaia G (mag)            & $6.413 \pm 0.003$     &1 \\
$B_{T}$ (mag)           & $7.235 \pm 0.066$     &10 \\
$V_{T}$ (mag)           & $6.597 \pm 0.010$     &10 \\
$J$ (mag)               & $5.486 \pm 0.034$     &11 \\
$H$ (mag)               & $5.248 \pm 0.047$     &11 \\
$K_{s}$ (mag)           & $5.133 \pm 0.023$     &11 \\
WISE \textit{W1} (mag)  & $5.080\pm0.201$   &12\\
\hline
\end{tabular} \\
{\sc Note:} References abbreviated as 
(1) \citet{GaiaEDR3_2020}; (2) \citet{Soubiran_2018}; (3) \citet{Soubiran_2018}  (4) \citet{Aguilera_2018};  (5) \citet{Wright_2004}; (6) \citet{Murgas_2013}; (7) \citet{Brewer_2016}; (8) \citet{Gomes_2021}; (9) \citet{Voges_1999}; (10) \citet{Hog_2000}; (11) \citet{Cutri_2003}; (12) \citet{Cutri_2014}; (13) This work
\end{table}

Using chromospheric activity as a proxy activity-age indicator, it is possible to attain an age estimate for \Targetname~A. The chromospheric activity index $\log(R'_{\rm HK})$, measured by the S-index \citep{Vaughan_1980,Duncan_1991}, is calculated by determining the ratio of the flux in a narrow bandpass centered on the Ca\,{\sc ii} H and K lines to the flux in two adjacent continuum bandpasses. \Targetname\ has an S-index ranging from 0.127-0.139 in the \citet{Pace_2013} catalog. Literature measurements report $\log(R'_{\rm HK})$ values of $-5.12\pm0.07$ dex \citep{Wright_2004,Murgas_2013,Brewer_2016,Gomes_2021}, which favor an intermediate age for the system, spanning the range of 2.4-3.57 Gyr \citep{Marsakov_1995,Valenti_2005,Takeda_2007,Holmberg_2009,Delgado_2015,Luck_2017,Yee_2017,Delgado_2019}. We undertake an independent measurement of the stellar activity age using the Bayesian age-dating method outlined in \citet{Brandt_2014} and applied in \citet{Li_2021}. This approach incorporates the X-ray ($R_{\rm X}$) and chromospheric activity ($ R'_{\rm HK}$) indicators, along with the optional inclusion of rotation period to estimate ages using the calibrated activity-age relation provided by \citet{Mamajek_2008}. We derive an activity-based age of $\ge$3.4 Gyr at a 68$\%$ confidence level given a chromospheric index value of $\log(R'_{\rm HK}) = -5.12\pm0.07$ and an X-ray index of $R_{\rm X}<-5.32$ from \citet{Voges_1999}. Figure~\ref{age_posterior} shows the activity-based posterior probability distribution for the age of \Targetname~A. Our result only offers a lower estimate of the stellar age and does not impose an upper limit. For comparison, BAFFLES \citep{Stanford-Moore_2020} yields age lower limits of $>6$ Gyr and $>3$ Gyr at $68\%$ and $95\%$ confidence levels, respectively, based on a chromospheric index of $\log(R'_{\rm HK}) \approx -5$ dex and a B-V color of $B-V = 0.58 \pm 0.002$ \citep{Hipparcos_1997}. Note that these lower limits from BAFFLES are marginally underestimated because we used an approximated chromospheric index due to the calibration range limited to -5 dex in \citet{Stanford-Moore_2020}. In all scenarios, our result is consistent with the prevailing literature measurements supporting an intermediate age for the star, most likely slightly younger than the Sun \citep{Silva_2020, Delgado_2019, Llorente_2021, Palla_2022, Luck_2017}.

The mass of \Targetname~A has also been the subject of various studies. \citet{Bochanski_2018} matched the Gaia observations with data from the 2MASS and Wide-Field Infrared Survey Explorer catalogs, and applied MIST isochrones to derive an estimate of $1.46\pm0.05 \Msun$ for \Targetname~A among a sample of co-moving stars. \citet{Anders_2022} obtained a mass of $1.27\pm0.19 \Msun$ by melding photometric and astrometric data from Gaia EDR3 with stellar evolutionary models. \citet{Gomes_2021} analyzed HARPS spectra and determined a mass of $1.426\pm0.017 \Msun$, while \citet{Paegert_2021} utilized Bayesian inference with Gaia parallax data, obtaining a mass of $1.12\pm0.15 \Msun$ for the TESS input catalog. While the literature currently contains a range of mass estimates, astroseismology with TESS could enable a more accurate determination of the host star's mass. We independently determined the stellar mass with the PARSEC isochrone stellar evolution models \citep{Bressan_2012} following \citet{Li_2021}. We derive a mass of $\rm 1.41\pm 0.05 \Msun$ that is consistent with the literature. Due to the diverse  mass estimates found in the literature, we not only consider our mass estimate of $1.41\pm0.05 \Msun$ as a stellar mass prior in Section~\ref{sec:orbit_constrain}, but also explore a broader stellar mass prior of $1.41 \pm 0.15 \Msun$ to encompass the range of values reported in previous studies.

\section{The Predicted Companion from RV and Absolute Astrometry}
\label{sec:prediction}

In 2021, we initiated a pilot direct imaging survey program at the Keck Observatory to find companions around accelerating stars. \citet{Brandt_2019} has shown that stars demonstrating significant ($\ge 3 \sigma$) disparities between \Hipparcos and \Gaia proper motions in the cross-calibrated \Hipparcos-\Gaia Catalog of Accelerations (HGCA) \citep{Brandt_2018,Brandt_2021_hgca} point to potential hosts of unseen companions. Our target selection strategy uses HGCA as a tool to identify stars with notable proper motion anomalies between \Hipparcos\,and \Gaia. In addition, stars with known radial velocities allow us to utilize both HGCA data and the RVs to jointly fit their orbits using the MCMC orbit code $\orvara$ \citet{Brandt_2021} before initiating observations. This precursor information garnered from the RV and HGCA astrometry orbital fit enables us to predict potential locations of companions for follow-up imaging efforts. The combination of an HGCA-based target vetting strategy and $\orvara$ location predictions represents a novel approach to improve high-contrast imaging yields, as demonstrated by the success of several other highly productive imaging surveys such as \citet{Fontanive_2019,Currie_2021,Bowler_2021, Bonavita_2022,Franson_2022b,Rickman_2022}. 

\Targetname~B was selected for our pilot program with a vetting technique similar to that discussed in \cite{Franson_2023b}, which involves calculating joint probability maps in separation and companion mass for stars with significant low-amplitude HGCA accelerations. The probability maps predict allowed and disallowed zones for potential companions around accelerating host stars based on their accelerations in the HGCA, helping evaluate the potential discovery space. We also conduct preliminary MCMC orbital retrieval using $\orvara$ and RV+HGCA data to predict the location of potential companions at specified observation epochs for subsequent direct imaging observations.

\subsection{Absolute Astrometry}

\begin{table*}
\centering
\caption{HGCA absolute astrometry for \Targetname~A.}
\label{table:hgca_absAst}
\begin{tabular}{lccc}
	    \hline
Parameter & Hipparcos & Hipparcos-Gaia & Gaia EDR3 \\
\hline
$\mu_{\alpha*}$ (mas\,yr$^{-1}$)   &$-35.748 \pm0.492$ & $-33.565 \pm0.016$ & $-33.726 \pm 0.025$\\
$\mu_{\delta}$ (mas\,yr$^{-1}$)    &$-129.582 \pm 0.386$ & $-127.543\pm 0.013$ & $-125.768 \pm0.025$\\
${\rm corr}(\mu_{\alpha*},\mu_{\delta})$     &0.237 &0.177 & $-0.086$ \\
$t_{\alpha}$ (Jyr)    &1991.33   & -- & 2015.87 \\
$t_{\delta}$ (Jyr)    &1991.48 &-- & 2015.85 \\\hline
% $\chi^{2}$  & 4055.8037\\
\end{tabular} \\
{\sc Note:} The $\chi^2$ value for a model of constant proper motion (Hipparcos-Gaia \\ and Gaia proper motions are equal) is 4056 with two degrees of freedom.
\end{table*}

The absolute astrometry of \Targetname A is taken from the Gaia EDR3 version of the Hipparcos-Gaia Catalog of Accelerations (HGCA, \citet{Brandt_2021_hgca}). The HGCA is a cross-calibration of the Hipparcos and Gaia EDR3 catalogues onto a common reference frame with well-characterized uncertainties. The HGCA compares the 25-year Hipparcos-Gaia scaled positional difference, the Hipparcos proper motion (near 1991.25), and the Gaia EDR3 proper motion (near 2016.0).  Different proper motions suggest reflex motion of the star due to the tug of a massive companion. 

\Targetname~A was selected for our program as it experiences proper-motion differences of $\approx$64$\sigma$ in the HGCA, suggesting the existence of a massive companion. Table~\ref{table:hgca_absAst} lists \Targetname~A's absolute astrometry from the HGCA, including the Hipparcos proper motion, the Gaia DR3 proper motion, and a joint Hipparcos-Gaia positional difference divided by the temporal baseline between the two missions. The Renormalized Unit Weight Error (RUWE) metric is used to assess the quality of astrometric solutions provided by Gaia, with values close to 1 indicating a good fit between the observations and the expected behavior of a single star. The RUWE for \Targetname\ in Gaia is 1.02 \citep{GaiaEDR3_2020}, well below the threshold of 1.4 at which the Gaia pipeline attempts a non-single star solution. This indicates that the Gaia observations for \Targetname\ are consistent with a satisfactory single-star solution: there is little orbital curvature within the Gaia data.

\begin{figure}
    \centering
\includegraphics[width=0.5\textwidth]{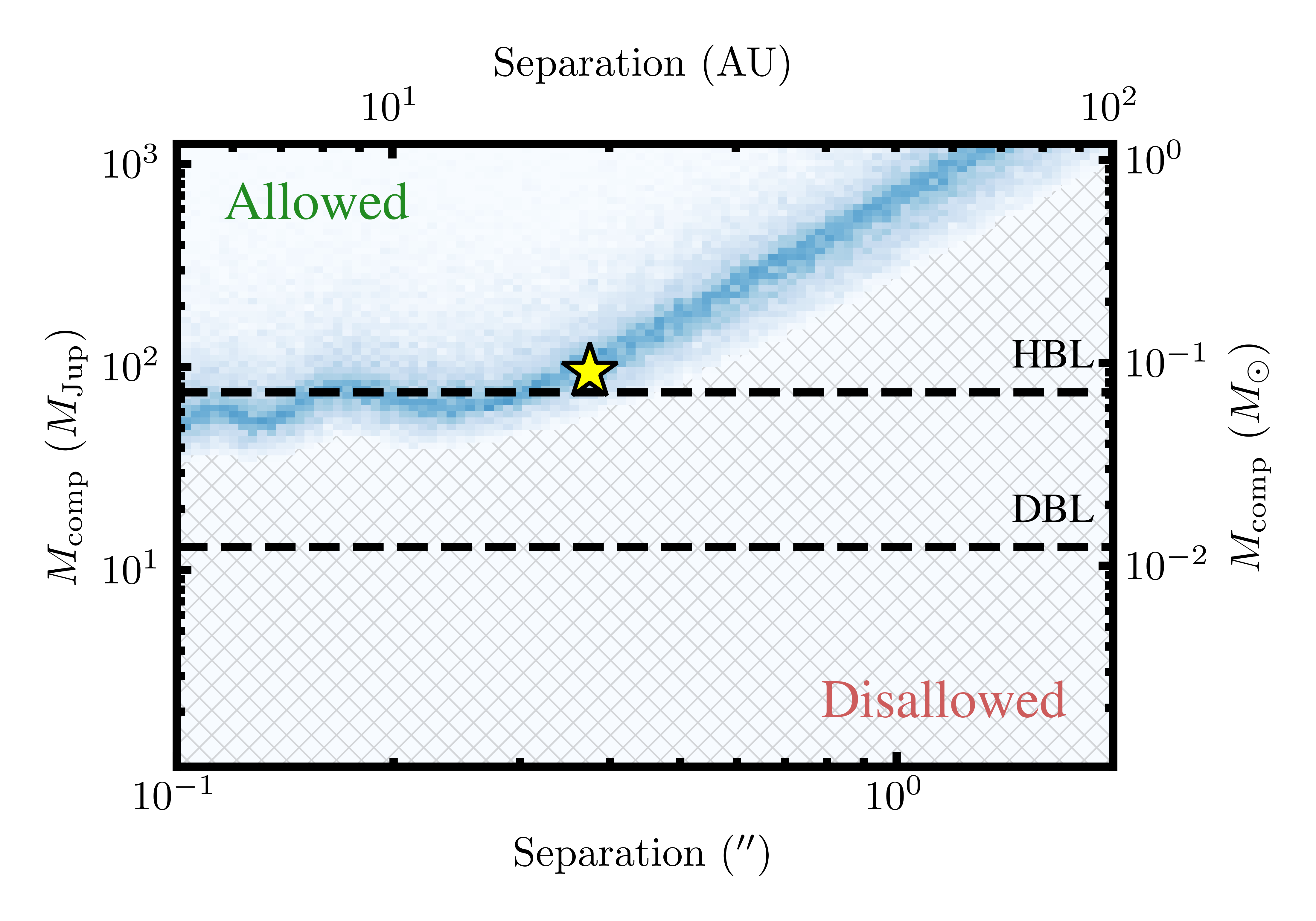}
    \caption{Predicted mass of \Targetname~B as a function of projected separation based on the \Hipparcos-\Gaia proper motion anomaly of its host star \Targetname~A. The color gradient indicates the relative probability of agreement between the predicted astrometric mass and the dynamical mass, with companions below the blue curve unable to account for the observed change in proper motion. \Targetname's astrometric acceleration points to a substellar companion within 0.2'' or a more distant star. Dotted lines mark the Hydrogen Burning Limit (HBL) of $\sim$ 75 $\Mjup$ and the Deuterium-burning limit (DBL) of $\sim$ 13 $\Mjup$. The yellow star represents \Targetname~B's predicted dynamical mass ($95_{-19}^{+25} \Mjup$), eccentricity ($\sim$ 0.26) and separation ($\sim$ 0.4'') based on the RV+HGCA $\orvara$ fit in Section~\ref{predicted_location_RVHGCA}, predating direct imaging observations.
  \label{hgca_prediction}}
\end{figure}

\begin{figure*}
    \centering
    \begin{minipage}[t]{0.5\textwidth}
        \centering
\includegraphics[height=\textwidth]{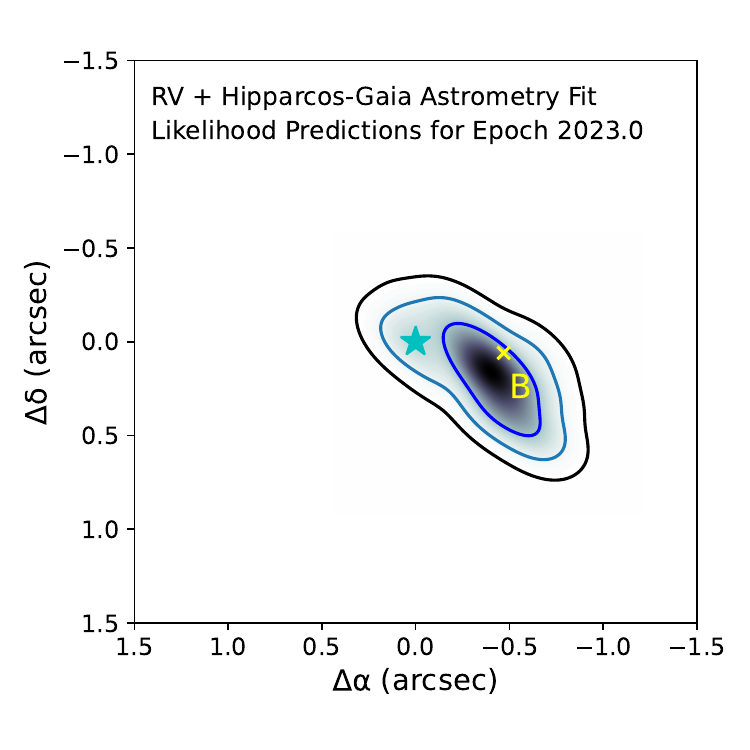}
    \end{minipage}\hfill
    \begin{minipage}[t]{0.45\textwidth}
        \centering
        \begin{adjustbox}{raise=+1.53em}
        \includegraphics[height=\textwidth]{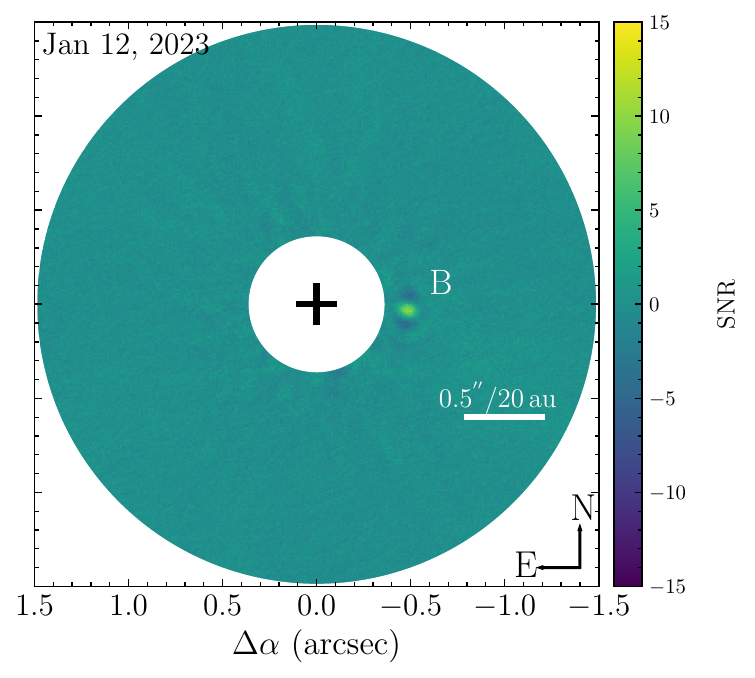}
        \end{adjustbox}
    \end{minipage}
    \caption{(Left) The blue probability contours depict the predicted position (1, 2, and $3\sigma$) of \Targetname~B relative to its host star (cyan). The companion's location was constrained by fitting archival radial velocities from HARPS and HIRES, and Hipparcos-Gaia absolute astrometry with the $\orvara$ orbit fitting package \citet{Brandt_2021}. The yellow cross shows where it was imaged. (Right) PCA PSF subtracted image with nine principle components showing the detection of the companion within the edge of its predicted 1-$\sigma$ likelihood contour. The SNR of the detection is 12.5. \label{Annular_Psf}}
\end{figure*}

\subsection{Radial Velocity}

The RV of the system was monitored by both HARPS and HIRES. The HARPS and HIRES radial velocity data do not cover the full orbital period, but suggest a shallow trend for \Targetname~A, indicating the presence of a distant companion. The HIRES RVs are acquired from the High Resolution Echelle Spectrograph (HIRES) \citep{Vogt_1993} positioned at the right Nasmyth focus of the Keck-1 telescope. 
These RVs encompass observations both before and after the 2004 CCD upgrade, collectively demonstrating an rms value of 26.3 m/s, with a median uncertainty of 1.9 m/s for each individual measurement. 
The High Accuracy Radial Velocity Planet Searcher (HARPS) spectrograph was installed in 2002 on the ESO's 3.6m telescope at La Silla Observatory in Chile. The HARPS RVs exhibit an rms value of 13.0 m/s and a median uncertainty on each individual measurement of 1.46 m/s. We fetch the data from the HARPS-RVBANK's \citep{Trifonov_2020} ``DRVmlcnzp" column, which corrects for zero-point variations, intra-night RV drift, and an absolute RV offset attributed to the upgrade of the HARPS fibers in 2015. In our orbit analysis, we treat the RV datasets from both the HAPRS and HIRES before and after the fiber upgrades as separate instruments with distinct RV zero points. In total, we consider four instruments and four RV datasets.

\subsection{Predicted Location}
\label{predicted_location_RVHGCA}

First, we use RVs and absolute astrometry as described above to perform an orbital fit to the \Targetname system. In addition to modeling the orbital parameters, $\orvara$ allows for the generation of 1-, 2-, and 3-$\sigma$ likelihood contours that illustrate the potential coordinates of the companion relative to its host star at any given epoch. The initial results from the RV and absolute astrometry fit indicate the presence of a massive stellar companion of $95^{+25}_{-19}\,\Mjup$ orbiting \Targetname~A at a distance of $18.8^{+10}_{-4.8}$ AU. We show the predicted 3-$\sigma$ contours outlining the potential positions of the companion in the left panel of Figure~\ref{Annular_Psf}. The RVs and absolute astrometry point to a companion west of the star, with mean predicted separations in RA and Dec at 2023.0 of
$-390 \pm160 $ mas and $180\pm 150$ mas, respectively. 

We also compute joint probability maps in separation and companion mass for \Targetname~A based on its HGCA astrometric accelerations following \citep{Franson_2023}. This involves the generation of 100 circular orbits with random orientations for each grid point in semi-major axis-companion mass space. The resultant acceleration distribution is then juxtaposed with the star's average acceleration and uncertainty, employing the K-S statistic \citep{Kopytova_2016}. The mass prediction based on \Targetname~A's acceleration is showcased in Figure~\ref{hgca_prediction}, suggesting the existence of either a brown dwarf within 0.2'' or a stellar companion beyond 0.2''. Our predicted companion, represented by the yellow star in the RV+HGCA only fit, indicates a stellar companion with a dynamical mass of $95^{+25}_{-19}\,\Mjup$ at a separation of $\sim 0.4''$. The preliminary RV+HGCA fit places the stellar companion within the allowed zone, hinting at the possibility of discovery. However, the orbit of the companion is subject to variability as direct imaging observations are incorporated. In the next section, we describe our thermal infrared imaging discovery of the predicted companion \Targetname~B.

\section{Keck/NIRC2 Observation and Data Reduction}
\label{sec:observation}

We observed the \Targetname system with the thermal Near-Infrared Camera 2 (NIRC2) at the Keck Observatory in the $L'$-band (central wavelength 3.8 $\mu m$) on UT 2023 Jan 11. We use the  narrow camera with a pixel plate scale of 9.971$\pm$0.004 mas/pixel \citep{Service_2016} and a 512$\times$512 pixel subarray. The NIRC2 data were taken behind the Vector Vortex Coronagraph (VVC; \citealt{Serabyn_2017}) with a vortex phase mask using natural guide star AO and the visible-light Shack-Hartmann wave-front senser. The seeing conditions for the night were roughly photometric, with an average value of $1.\!\!''0\pm0.\!\!''3$ as seen by the differential image motion monitor (DIMM). The observations involve sequences of 20-30 science frames using the Quadrant Analysis of Coronagraphic Images for Tip-Tilt Sensing (QACITS) algorithm. This algorithm applies small tip-tilt corrections to re-center the star after each exposure, achieving milliarcsecond stability \citep{Huby_2015,Huby_2017}. We also incorporate a 4.5 mas QACITS centering uncertainty from \citet{Huby_2017} to account for the average pointing accuracy from the QACITS controller. This centering uncertainty is divided by the separation measurement and added in quadrature to the position angle uncertainty. Each sequence also includes an off-axis unsaturated frame for flux calibration and sky-background frames for both the science images and the point-spread-function (PSF) images. We captured a total of 165 science images, excluding 9 short pointing optimization frames. Each frame consisted of 90 coadds with 0.3s exposures to for optimal readout efficiency. The total integration time was 4590s (76.5 minutes), accompanied by a field rotation of $35^{^\circ}$. 

We process our imaging sequence using Angular Differential Imaging (ADI) with the Vortex Image Processing ($\tt VIP$) package \citep{Gomez_2017}. $\tt VIP$ is specifically designed for processing vortex coronagraphic observations. We first flat field and dark subtract our science images. Then we employ the $\texttt{lacosmic}$ Python package to remove cosmic rays \citep{vanDokkum_2001}, and we correct for geometric distortion using the solutions provided by \citet{Service_2016} for the narrow-field mode of the NIRC2 camera. To enhance the signal-to-noise ratio (SNR), we employ $\VIP$ to model and subtract the sky background in both the science frames and the off-axis flux-calibration frames. Precise centering of the images is crucial in ADI reduction to ensure proper alignment of the vortex with the center of the field of view. We fit a 2D Gaussian to the vortex's core halo in each image while determining the centroid. For each time slice, we cut a subimage centered around the centroid we have determined so that the images are aligned properly at the mas level between frames. The alignment uncertainty is gauged by computing the standard deviation of the centroid from the image center. The total alignment error, including QACITS centroiding uncertainty, averages around 5.0 mas.

\begin{figure*}
    \centering    \includegraphics[width=0.7\textwidth]{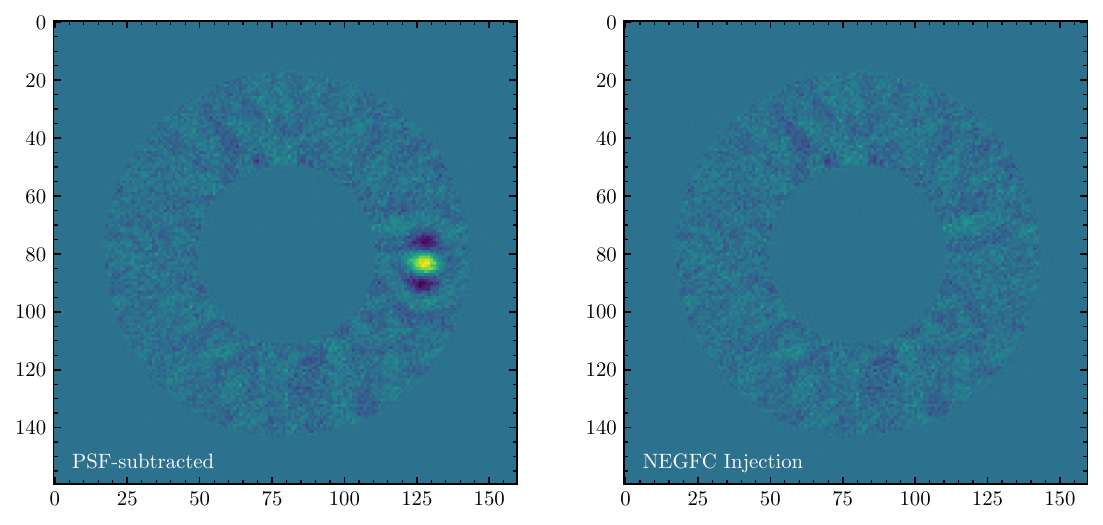}
    \caption{PSF-subtracted image of \Targetname~B (left) and the same image after introducing negative PSF templates with optimal separations, position angles, and flux ratios (right). The negative companion injection successfully eliminates companion signals in the latter image. \label{negfc_injection}}
\end{figure*}

$\VIP$ offers PCA-based postprocessing algorithms \citep{Soummer_2012, Amara_2012} for stellar point spread function (PSF) subtraction. The PSF is obtained by subtracting the
sky-background frame from the PSF image with the star off the vortex. By fitting a 2D Gaussian profile to the host-star's sky-subtracted PSF, we measure a mean FWHM value of 7.89 pixels ($0.079''$). Here we use the same procedure outlined in \citet{Li_2023} to do PSF subtraction. Unlike simple median combination of ADI, PCA algorithms enhance the contrast by modeling the background including leaked starlight, static and quasi-static speckles. This is achieved by projecting each image onto the first $n$ principal components, obtained through singular value decomposition of a 2D matrix created from the observed sequence of images. This approach removes the stellar PSF from each image in \Targetname~A's data cube by subtracting principal components from individual annular regions. This highlights fainter and more extended structures like planets or disks. To reduce self-subtraction, we employ a rotation gap/PA threshold and radial scaling criterion \citep{Lafreniere_2007} in annular patches, discarding frames with inadequate rotation. Specifically, using $\VIP$, we find that setting a PA threshold of 0.3 FWHM, or $\delta > 0.3 \lambda/D$, where $\lambda$ is the observing wavelength and D is the diameter of the telescope's primary mirror, best ensures that the companion is not inadvertently subtracted during the image processing. We also tune the optimal number of components that would optimize the SNR by running PSF subtractions from 1 to 30 components, measuring the resultant companion SNR each time in the reduced images. SNR is determined using the method outlined in \citet{Mawet_2014}, which incorporates a penalty at small separations to account for the limited number of resolution elements. We found that the highest SNR is produced by using 9 components. We detected the companion with a signal-to-noise ratio (SNR) of $12.5$ using annular PCA. Figure~\ref{Annular_Psf} shows the detection of a point source with PCA subtraction, which we refer to as \Targetname~B, roughly 47.2 pixels or $0.\!\!''471$ from the star. \Targetname~B is the only object identified by NIRC2 within $1''$. We visually inspect the Standardized Trajectory Intensity Map (STIM; \citet{Pairet_2019}), which did not show any presence of other signals.

The negative fake companion (NEGFC) technique \citep{Marois_2010,Lagrange_2010,Wertz_2017} is often used in conjunction with principal component analysis (PCA) or other PSF subtraction methods to accurately extract the position and flux of point-like sources, particularly for faint companions close to bright stars. The use of NEGFC avoids biases in astrometry and photometry from partial and self-subtraction. The NEGFC technique works by injecting a negative-amplitude PSF template at the location of the companion, effectively removing the signal from the true companion in the final image. This allows for a more accurate measurement of the true companion flux and position. NEGFC has been shown to be more effective for faint companions closer to the primary star, which can be easily affected by residual speckles or other systematic artifacts. We show the PSF-subtracted signal alongside the result after NEGFC injection in Figure~\ref{negfc_injection}.

\begin{table}
\caption{Keck/NIRC2 relative astrometry and $L'$-band photometric measurements of the \Targetname~AB system} \label{tab:nirc2_results} 
\centering
\begin{tabular}{cc}
\hline
\multicolumn{2}{c}{\textbf{\Targetname~AB}}  \\ \hline
Instrument                  &Keck/NIRC2      \\
Filter                      &$L'$            \\
Date (UT)                   &2023-01-11      \\
Epoch (yr)                 &2023.03         \\
\hline
\multicolumn{2}{c}{\textbf{Relative Astrometry}}\\
\hline
Separation (mas)            &$473.3\pm 5.2$    \\
PA ($^{\circ}$)             &$277.89 \pm 0.14$ \\
\hline
\multicolumn{2}{c}{\textbf{Photometry}}\\
\hline
$\Delta L'$ (mag)           &  $9.78\pm0.06$    \\
$L'_{*}$ Flux (mag)         &  $5.11\pm0.05$  \\
$L'_{p}$ Flux (mag)         &  $14.89\pm0.06$ \\
\hline
\end{tabular} \\
{\sc Note:} The $L'$ magnitude for the host star is transformed from its 2MASS H band magnitude using relations in \citet{Bessell_1988}.
\end{table}

Using $\VIP$, we take the astrometric and photometric measurements from annular PCA as initial conditions for the iterative NEGFC calculations. First, we utilize the Nelder-Mead downhill simplex optimization algorithm \citep{Nelder_1965} to provide the initial guesses that would be used as a starting state in MCMC optimization. Next, we explore the 3D parameter space of the flux within the aperture and polar positions using the emcee affine-invariant Markov Chain Monte Carlo (MCMC) ensemble sampler introduced by \citet{Foreman-Mackey_2013}. We run 100 walkers, each taking 1000 steps, resulting in a total of $10^5$ steps. We use the auto-correlation time based criterion $N/\tau \geq a_c$ with $a_c = 50$ \citep{Christiaens_2021} to evaluate convergence. To ensure reliable results, we discard the first $30\%$ of each chain as burn-in. We run the fit multiple times to ensure consistent and identical astrometry and photometry results for each run. We further correct for throughput and the ratio of exposure times between stellar PSF frames (100 coadds with 0.008 integration time per coadd) and the science frames that contain the companion (90 coadds and 0.3s exposures). The errors on the astrometry are calculated by adding in quadrature the uncertianties from the MCMC NEGFC injection technique, our pre-processing alignment uncertainty of 0.5 mas which includes the QACITS centroiding uncertainty of 0.2 pixels. We correct for the position angle from the NEGFC reduction $\theta_{\mathrm{meas}}$ through 
\begin{equation}
\mathrm{\theta = \theta_{\mathrm{meas}} - (\mathrm{PARANG} + \mathrm{ROTPOSN} - \mathrm{INSTANGL} - \theta_{\mathrm{north}})}
\end{equation}
where PARANG is the parallactic angle offset, ROTPOSN is the rotator position of $4.\!\!^{\circ}43$, INSTANGL is the NIRC2 position angle zero-point of $0.\!\!^{\circ}7$, and $\theta_{\mathrm{north}} = 0.\!\!^{\circ}262\pm0.\!\!^{\circ}020$ \citep{Service_2016} is the angle we applied to rotate the frames counterclockwise in order for them to have a North-up and East-left orientation. We measure a corrected position angle of $277.\!\!^{\circ}89 \pm 0.\!\!^{\circ}14$ south-west from the primary star. Our final astrometry results based on the NEGFC approach are listed in listed in Table~\ref{tab:nirc2_results}. 

Given the lack of a multi-wavelength spectrum for \Targetname~B, our capacity to determine its spectral type is limited to deducing a bolometric luminosity from its $L'$ band brightness, followed by comparison to field brown dwarfs with known spectra. We infer a bolometric luminosity for \Targetname~B based on its $L'$ band magnitude using a method outlined in \citet{Li_2023}. First, we measure an $L'$ contrast between the star and the companion of $9.79\pm0.06$ mag. The $L'$ magnitude of the star is $5.11\pm0.05$ magnitudes, obtained through transformation from the 2MASS H band magnitude using a relation detailed in \citet{Bessell_1988}. This establishes the $L'$ mag for \Targetname~B as $L'_{p}$= $14.89\pm0.06$ mag, which corresponds to an absolute magnitude of $L'_{p}$ = $11.39\pm0.06$ mag based on the parallax. We transform this absolute $L'$ band magnitude into an absolute magnitude in the \textit{W1} band of $M_{W1}$=$12.20\pm0.07$ mag according to the relation derived by \citet{Franson_2023}. To retrieve the bolometric luminosity, we convert the absolute magnitude in the \textit{W1} band by fitting a fourth order polynomial to a segment of \citet{Filippazzo_2015} spectral energy distribution (SED) for a comprehensive sample of field dwarfs complied from legacy surveys. The uncertainty on the bolometric luminosity include the error in the $L'$ band contrast, the scatter in the $L'$ to \textit{W1} conversion by \citet{Franson_2023}, and the uncertainty in our fourth order polynomial fit converting \textit{W1} to bolometric luminosity. Overall error propagation resulted in a final bolometric luminosity of $\mathrm{log(L_{bol}/\Lsun)= -4.55 \pm0.08}$ dex for \Targetname~B. This luminosity implies that the companion resides in close proximity to the L/T transition.

\section{Orbital Constraints}
\label{sec:orbit_constrain}

\begin{figure*}
    \centering
    \includegraphics[width=0.49\textwidth]{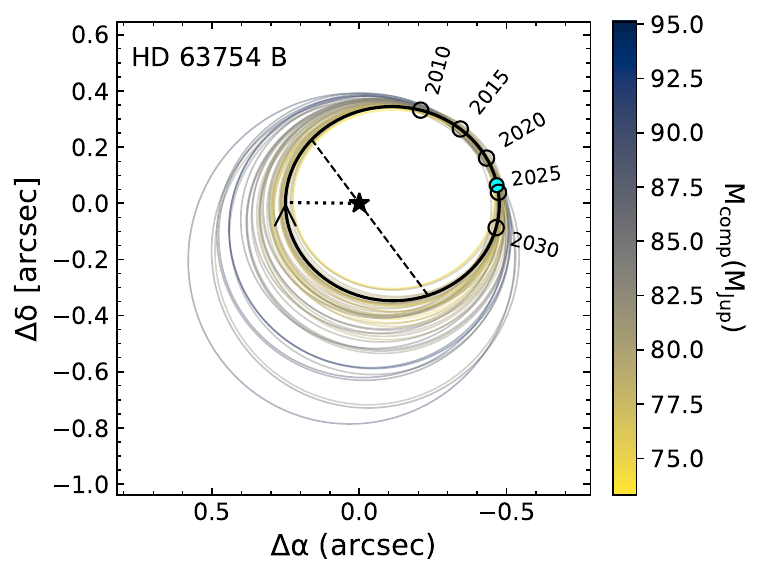}
    \includegraphics[width=0.43\textwidth]{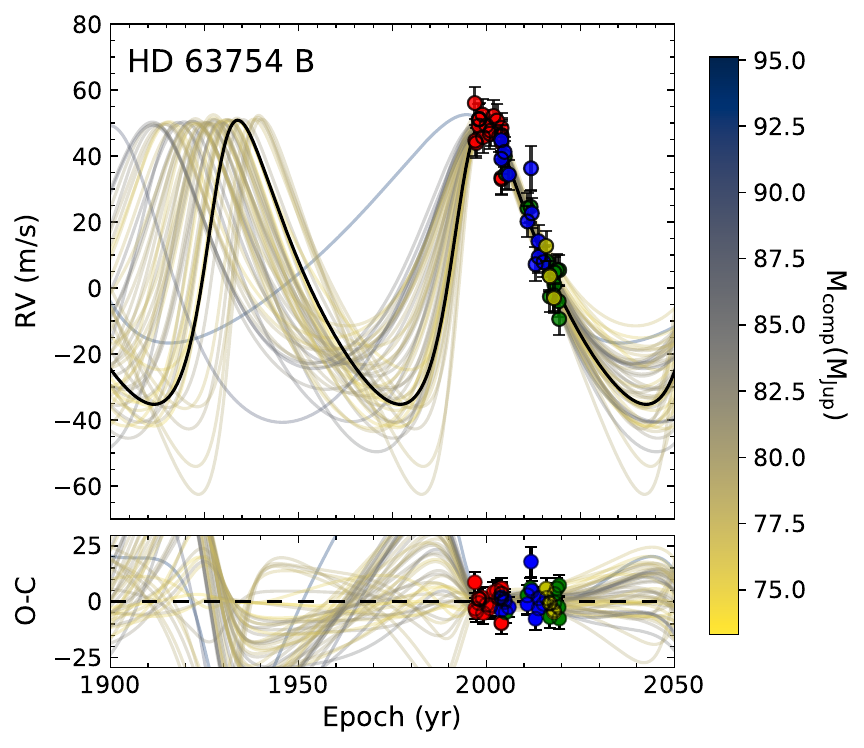}\quad
    \vspace*{+10mm}\hspace*{+5mm}
    \includegraphics[width=0.44\textwidth]{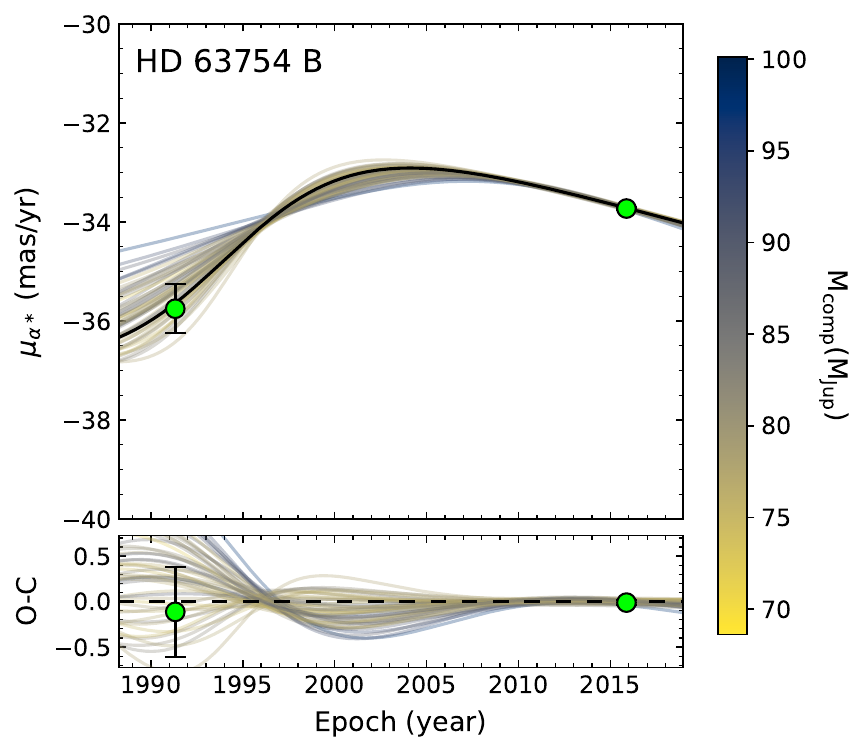}
    \includegraphics[width=0.44\textwidth]{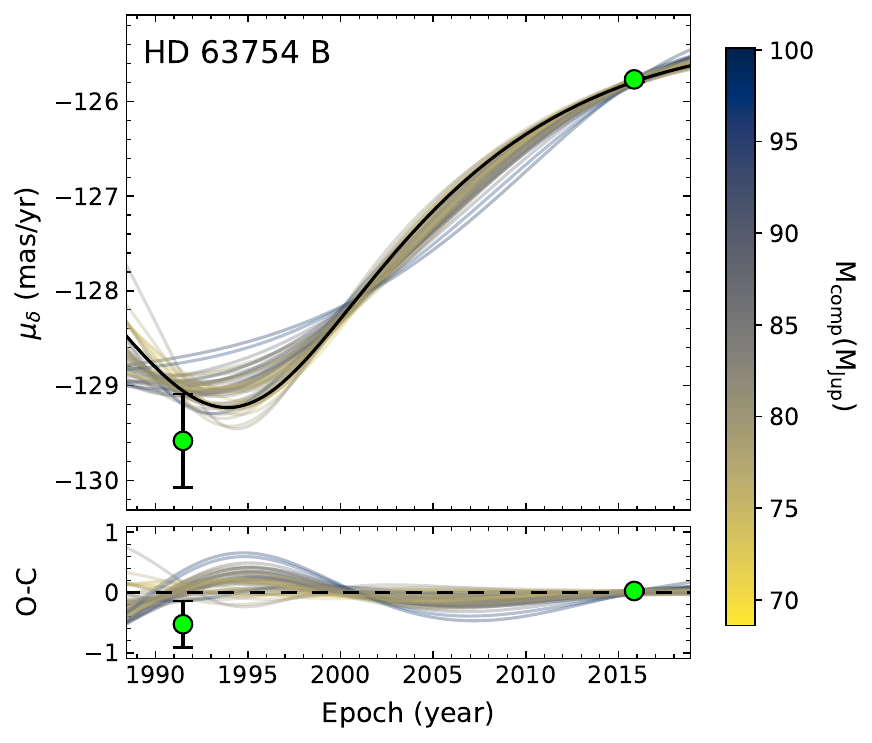}
    \caption{Orbit fit of the HD~63754~AB system, including the sky-projected orbit (top left), RVs (top right), and HGCA proper motions (bottom panels). The colorful curves represent 50 randomly drawn orbits from the MCMC chains, while the maximum-likelihood orbit is highlighted in black. The red, blue, yellow, and green data points in the RV datasets are from the HARPS pre-fiber update, HARPS post-fiber update, HIRES pre-fiber update, and HIRES post-fiber update, respectively. The green points in the lower panels represent the Hipparcos and Gaia EDR3 proper motions; the measured long-term proper motion (not shown) constrains the integral of the proper motion.  \label{fig:orbit_fit_plots}}
\end{figure*}

\renewcommand{\arraystretch}{1.2}
\begin{table*}
\caption{MCMC orbital fit results for \Targetname~B with the narrow Gaussian prior of 1.35$\pm 0.15\Msun$ on the host star's mass.}
\label{orbit_MCMC_result}
\begin{tabular}{ccccc}
\hline
\textbf{Parameter} & \textbf{Prior} & 
\textbf{Best Fit}& 
\textbf{68.3$\%$ CI}& 
\textbf{95.4$\%$ CI} \\ \hline
\multicolumn{5}{c}{Fitted Parameters}\\
\hline
$\sigma_{\rm Jit}$\,(m/s)                &1/$\sigma_{Jit}$       &4.41       &${4.41}_{-0.54}^{+0.64}$          &(3.413, 5.794)\\
$\rm M_{*}\,(\Msun)$                     &$N(0.72, 0.02)$        &1.40       &${1.40}_{-0.14}^{+0.14}$          &(1.122, 1.685)\\
$\rm M_{p}\,(\Msun)$                     &1/$M_{p}$              &81.9       &${81.9}_{-5.8}^{+6.4}$            &(70.663, 96.243)\\
$a$\,(AU)                                &1/a                    &19.9       &${19.9}_{-1.6}^{+2.7}$            &(17.251, 27.878)\\
$\sqrt{e}\sin \omega$                    &$U(-1,1)$              &-0.27      &${-0.27}_{-0.20}^{+0.34}$         &(-0.581, 0.432)\\
$\sqrt{e}\cos \omega$                    &$U(-1,1)$              &0.384      &${0.384}_{-0.12}^{+0.083}$        &(0.129, 0.529)\\
$i\,(^{\circ})$                          &$\sin i$               &174.81     &${174.81}_{-0.50}^{+0.48}$        &(173.781, 175.731)\\
$\Omega\,(^{\circ})$                     &$U(-180,180)$          &40.2       &${40.2}_{-7.1}^{+8.1}$            &(26.613, 57.906)\\
$\lambda_{\rm ref}\,(^{\circ})$          &$U(-180,180)$          &40.4       &${40.4}_{-8.4}^{+11}$             &(25.176, 64.963)\\
\hline
\multicolumn{5}{c}{Derived Parameters}\\
\hline
$\varpi$\,(mas)                          &--         &19.9319        &${19.9319}_{-0.0016}^{+0.0016}$       &(19.929, 19.935)\\
$P$\,(yr)                                &--         &73.4           &${73.4}_{-9.4}^{+16}$                 &(58.044, 123.531)\\
$\omega$\,$(^{\circ})$                   &--         &311            &${311}_{-271}^{+28}$                  &(4.301, 356.042)\\
$e$                                      &--         &0.260          &${0.260}_{-0.059}^{+0.065}$           &(0.148, 0.393)\\
$a$\,(mas)                               &--         &397            &${397}_{-33}^{+54}$                   &(344, 556)\\
$T_{0}$                                  &--         &2476422        &${2476422}_{-4083}^{+7530}$           &(2470000, 2496000)\\
\hline
\end{tabular} \\
{\sc Note:} The reference epoch is 2455197.5 JD. 
\end{table*}

We model the 3D orbit of \Targetname~B with the Bayesian orbit fitting code \texttt{orvara} \citep{Brandt_2021}, incorporating  HGCA absolute astrometry, HIRES and HARPS RVs, and single-epoch relative astrometry from our Keck/NIRC2 $L'$ band imaging. The \texttt{orvara} code employs parallel tempering MCMC (PT-MCMC) with the \texttt{emcee} ensemble sampler \citep{Foreman-Mackey_2013,Vousden_2015} to sample posterior orbital parameters. PT-MCMC involves running parallel chains at varying temperatures for precise sampling around the minimum $\chi^2$ within the parameter space; chains periodically exchange positions to enhance sampling efficiency. Our PT-MCMC process utilized 100 walkers, 30 temperatures, and a total of $10^{5}$ steps. \texttt{orvara} fits six orbital elements that fully describe a Keplerian orbit, plus two masses and an RV jitter term. The optimization speeds up parameter fitting by analytically handling auxiliary parameters like RV zero-points and parallax.

\begin{figure*}
    \centering
\includegraphics[width=\textwidth]{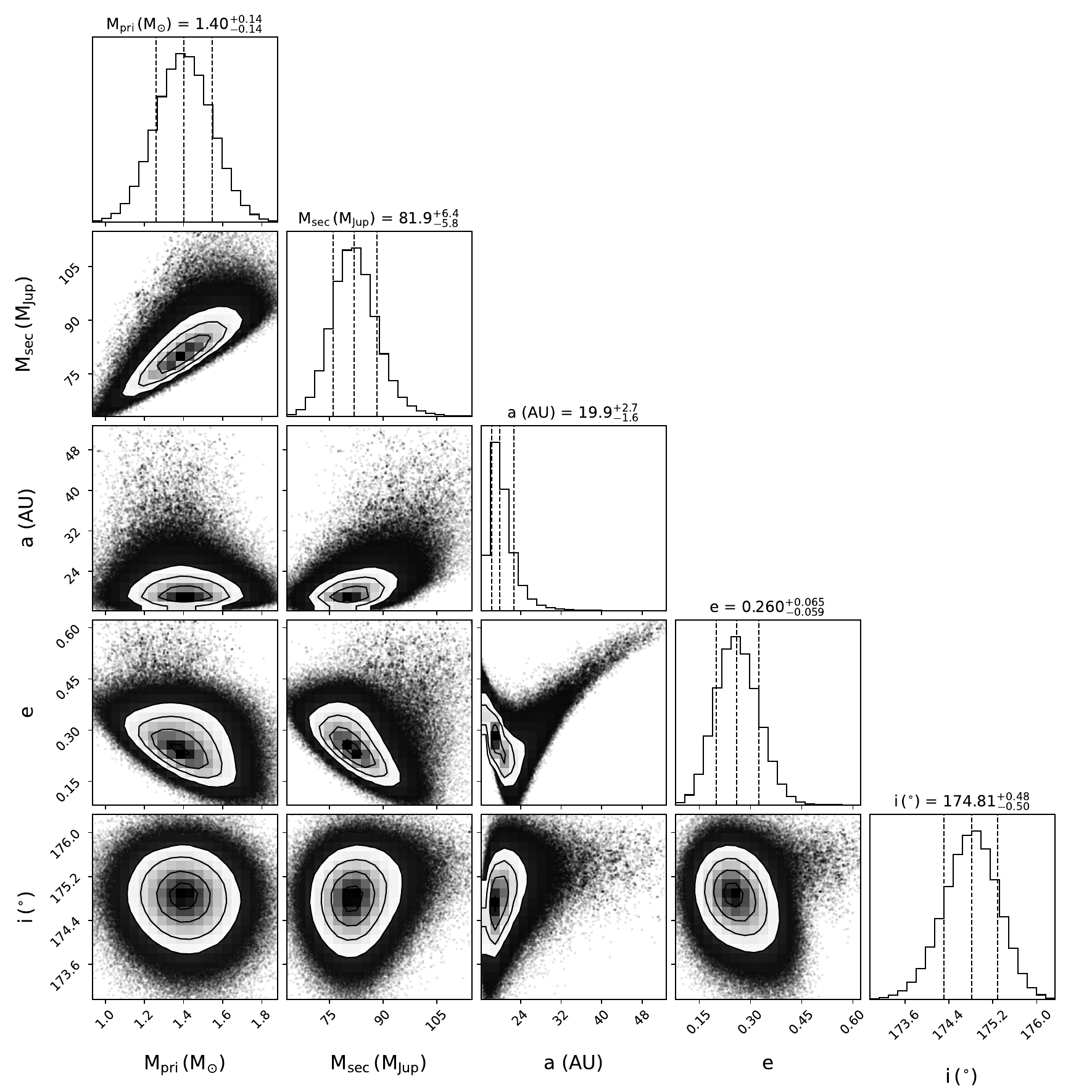}
    \caption{Posterior distributions using a Gaussian prior of $\rm 1.41\pm0.15 \Msun$ on the primary mass, both 1D and 2D, are presented for various orbital parameters of \Targetname~B. These distributions are derived from the analysis of radial velocity from HIRES and HARPS, relative astrometry obtained through Keck/NIRC2 direct imaging, and absolute astrometry from Hipparcos and Gaia data using $\orvara$ \citep{Brandt_2021}. The 1D posterior distributions are accompanied by confidence intervals at 15.85\%, 50.0\%, and 84.15\%, with the median $\pm$ 1$\sigma$ values indicated. Additionally, the 2D posterior distribution is visualized with contour levels corresponding to 1, 2, and 3$\sigma$.
\label{corner_orbfit}}
\end{figure*}

\begin{figure*}
    \centering
\includegraphics[width=\textwidth]{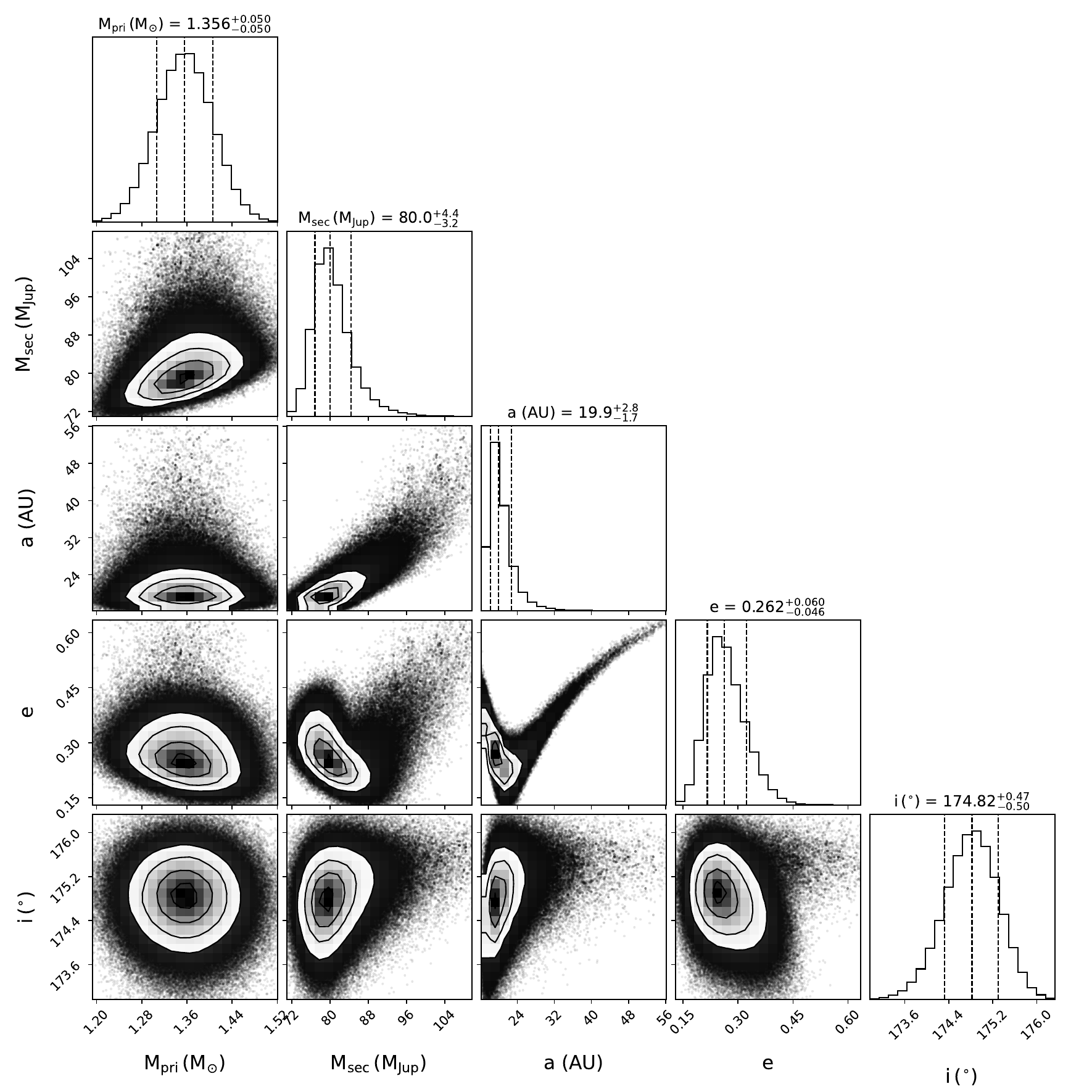}
\caption{Same as Figure~\ref{corner_orbfit} but using a narrower Gaussian prior of $\rm 1.41\pm0.05 \Msun$ on the primary mass in the MCMC orbital modeling. 
\label{corner_orbfit_narrow}}
\end{figure*}

We first impose an informative prior of $\rm 1.41\pm 0.05 \Msun$ on the mass of \Targetname~A, covering the range of literature measurements as discussed in Section~\ref{sec:properties}. We apply uninformative priors on all other fitted parameters: flat priors on the eccentricity, argument of periastron, mean longitude, and ascending node, and log-flat priors on the mass of the companion, semi-major axis and RV jitter, and geometric priors on the inclination. We assess convergence via visual inspection and discard the initial 40$\%$ of the chain as burn-in. We verify that variations in the burn-in threshold beyond 40$\%$ have marginal effects on resultant posteriors and consistently yield the same inferred parameters. Using this putative informative prior on the host star's mass, we obtain a dynamical mass of $80.0_{-3.2}^{+4.4} \Mjup$, a semi-major axis of $19.9_{-1.7}^{+2.8}$ AU, an eccentricity of $0.262_{-0.046}^{+0.060}$, and an inclination of $174.\!\!^{\circ}82_{-0.50}^{+0.47}$. To account for the uncertainty linked to the host star's mass, we also explore a broader prior of $\rm 1.41\pm 0.15 \Msun$. With this wider prior we measure a dynamical mass of ${81.9}_{-5.8}^{+6.4} \Mjup$ for the companion, a semi-major axis of ${19.9}_{-1.6}^{+2.7}$ AU, an eccentricity of ${0.260}_{-0.059}^{+0.065}$, and an almost face-on inclination of $174.\!\!^{\circ}81_{-0.50}^{+0.48}$. The corner plot illustrating the correlations among selected parameters for the informative prior of $\rm 1.41\pm 0.05 \Msun$ and the narrowed prior are shown in Figure~\ref{corner_orbfit} and Figure~\ref{corner_orbfit_narrow}, respectively.

Table~\ref{orbit_MCMC_result} enumerates the orbital parameters inferred from our MCMC posteriors. In Figure~\ref{fig:orbit_fit_plots}, we provide an illustrative selection of 50 orbits, randomly sampled from the posterior distributions, which showcase the relative orbit projected onto the sky plane, the RV fit, and the astrometric fit to the HGCA proper motions. The best fit orbit is represented by the black curve among the colorful randomly sampled orbits. The Hipparcos and Gaia proper motions are both formal good fits to the astrometric reflex motion of the star (a total $\chi^2 = 1.6$), with Gaia providing the more precise measurement of the two. The HARPS and HIRES combined RV measurements and our solitary astrometric data point cover only a segment of the orbit around aphelion, contributing to slightly better convergence in that orbital region. The lack of phase coverage by both RV and relative astrometry collectively contributes to the uncertainty in the orbital fit. We find a range of potential orbital solutions with fluctuations within a 6$\%$ uncertainty in the companion mass. 

\section{Discussion}
\label{sec:discussion}

\begin{figure*}
    \centering
    \hspace*{+0.45cm}
\includegraphics[height=0.29\textwidth]{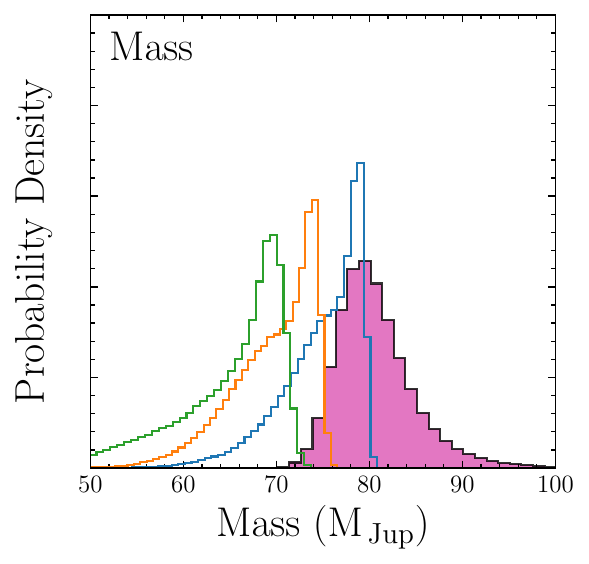}\quad
\includegraphics[height=0.29\textwidth]{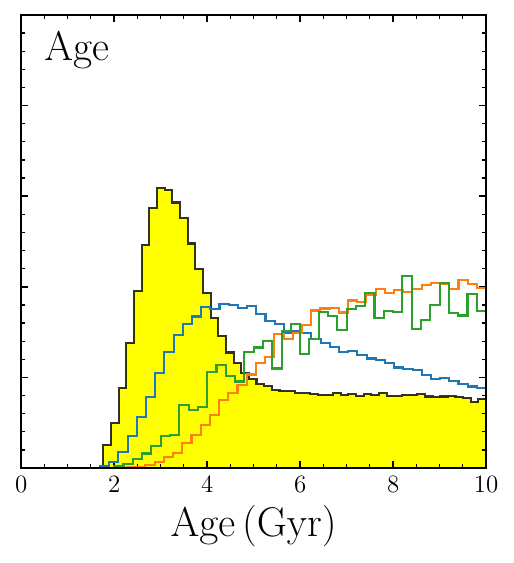} \hspace*{+2.5mm}
\includegraphics[height=0.29\textwidth]{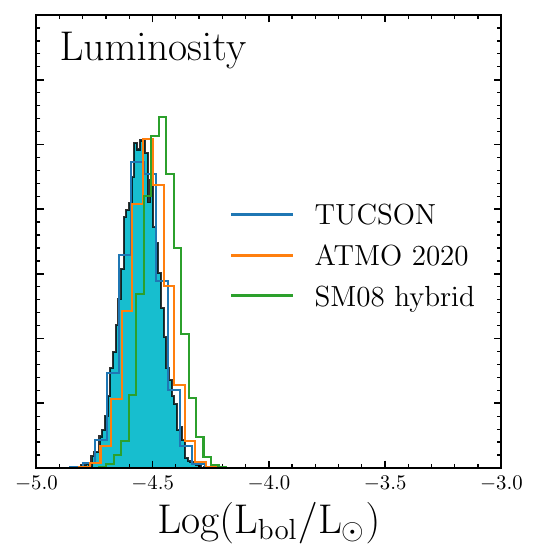}\quad
    \caption{Comparison between predicted masses and ages from evolutionary models and the dynamical mass and age of \Targetname~B. (Left) Posterior distribution of dynamical mass (filled magenta histogram) compared to model-derived mass posteriors for the inferred distributions of  $L_{\rm bol}$ and host star age. (Right) Posterior distribution of age sampled with evolutionary models from the inferred mass and luminosity. The filled yellow distribution represents the posterior from our Bayesian isochrone age-dating model. With a flat prior on age, all evolutionary models favor moderate to older ages for the system; this is due to the high measured mass and low luminosity. \label{mass_comparison}}
\end{figure*}

\begin{figure*}
    \centering
\includegraphics[height=0.7\textwidth]{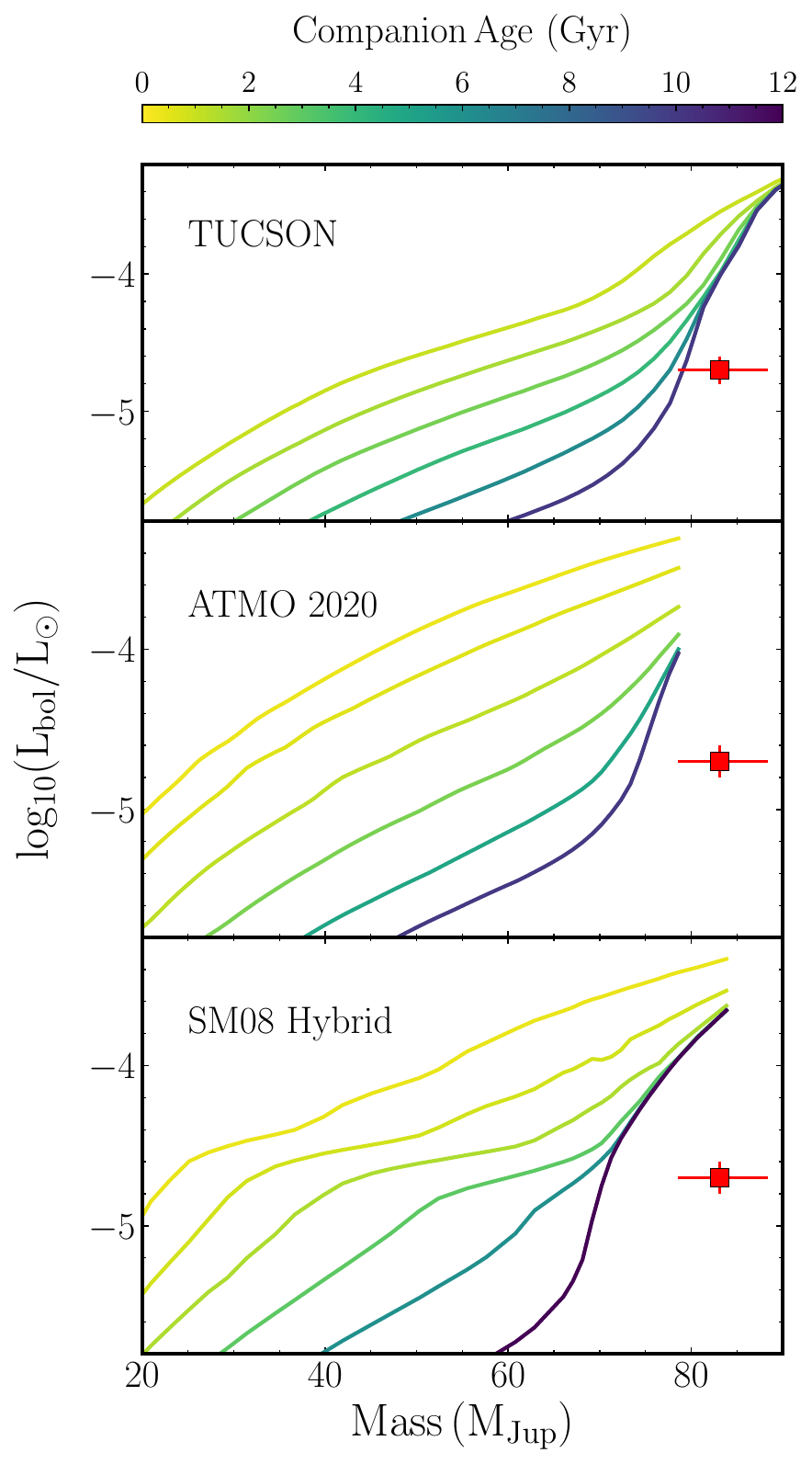}\quad \hspace*{-3mm}
\includegraphics[height=0.7\textwidth]{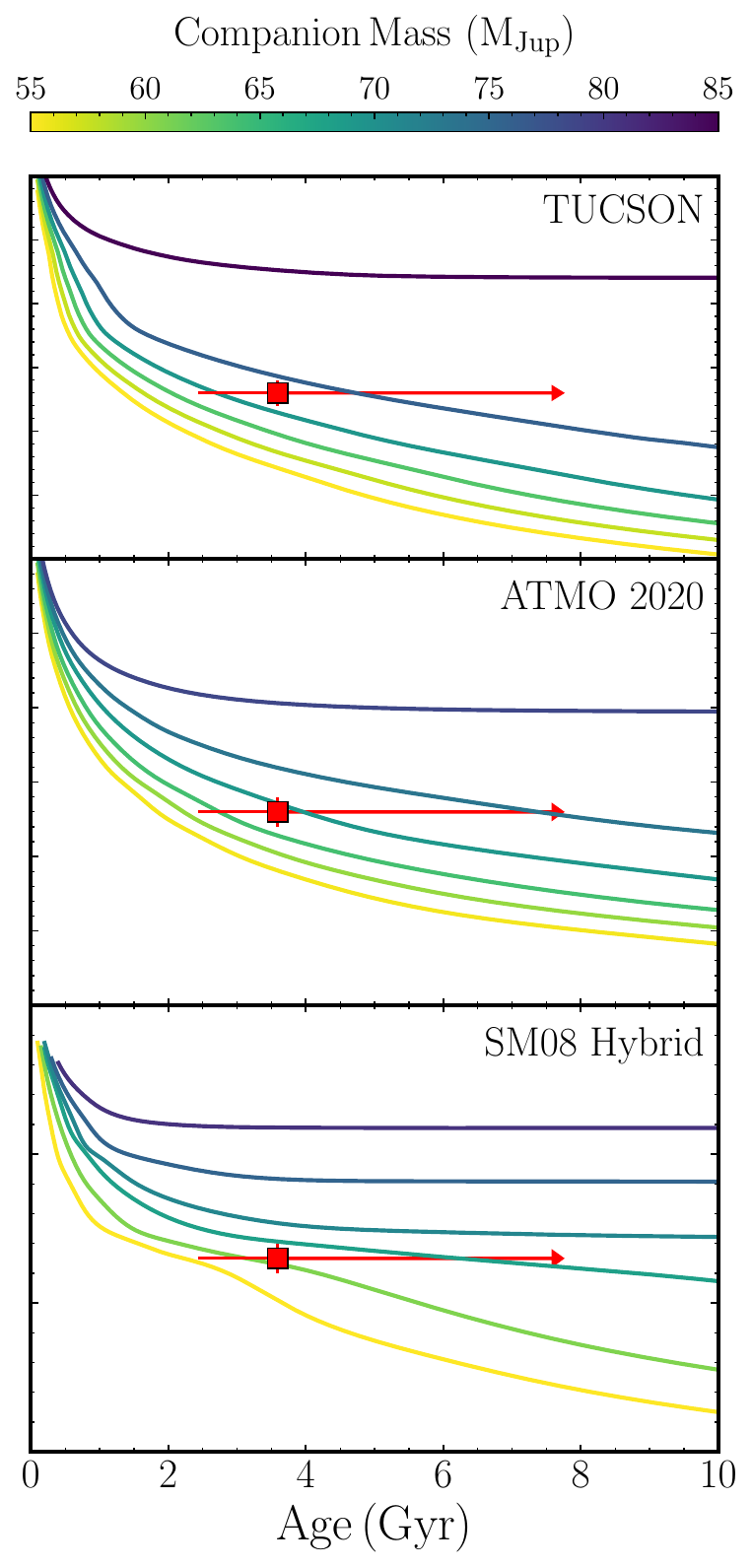}\quad
    \caption{Comparison of the dynamical mass, luminosity and age of \Targetname~B with the predicted masses and ages from three evolutionary models. Left panels show the age-luminosity plot and the iso-mass evolutionary grids from the three models. Right panels present the iso-age evolutionary grids. The grids are color-coded by the companion age and mass, respectively. The observed properties for \Targetname~B are indicated by the red square markers with errorbars. The overestimation of the dynamical mass and/or underestimation of bolometric luminosity could explain why the red data points fall outside the model grids in the left panels. \label{evol_grid}}
\end{figure*}

\begin{table*}
\caption{Inferred parameters from substellar evolutionary models vs. measured parameters for the companion. \label{model_derived_params}}  
\begin{tabular}{ccccc} 
\hline
\textbf{Property} & \textbf{TUCSON}                      & \textbf{ATMO2020}                    & \textbf{SM08 Hybrid}                 & Measurement \\ \hline
Mass ($\Mjup$)    &$75.0^{+5.3}_{-3.0} \,(1.5\sigma)$    &$69.8^{+5.6}_{-3.3} \,(2.8\sigma)$    &$66.1^{+8.1}_{-3.2}\,(3.9\sigma)$     &$81.9^{+5.8}_{-6.4}$  \\
Age (Gyr)   %\tablenotemark{c}  
&$4.7^{+2.9}_{-2.1}\,(1.8\sigma)$
&$6.4^{+4.3}_{-3.1}\,(2.7\sigma)$    &$6.0^{+3.6}_{-2.4}(2.5\sigma)$ 
&$\ge3.4$  \\
log($L_{\rm bol}/L_{\odot}$) &$-4.57^{+0.08}_{-0.08} \,(0.03\sigma)$ &$-4.53^{+0.08}_{-0.08} \,(0.02\sigma)$  &$-4.48^{+0.07}_{-0.08} \,(0.2\sigma)$                  &$-4.55\pm0.08$        \\
$T_{\rm eff}$ (K)   %\tablenotemark{d}   
&$1381\pm173$ 
&$1413\pm177$ &$1344\pm71$ &--                     \\
$\log(g)$ ($\rm cm\,s^{-2}$) %\tablenotemark{d}    
&$5.30\pm0.02$     &$5.43\pm0.02$     &$5.34\pm0.06$     &--                     \\
Radius ($R_{\rm Jup}$)%\tablenotemark{d} 
&$0.90\pm0.02$   &$0.86\pm0.02$   &$0.95\pm0.05$  &--      
%check the uncertianties in logg and radius, an order of magnitude off
\\ \hline
\end{tabular} \\
{\sc Note:} Masses were determined using measured $\log(L{\rm bol})$ and age, $\log(L_{\rm bol})$ was derived from measured mass and age, and age was estimated from measured mass and $\log(L_{\rm bol})$. Model-derived radii were based on the measured age and mass, while $\rm T_{eff}$ and log g were determined using model radii and measured mass. 
%\tablenotetext{b}{For each model, we use the measured mass and age posteriors to compute the bolometric luminosity.}

%\tablenotetext{c}{We use measured $\log(L_{\rm bol})$ and posterior mass distributions to estimate the age. }

%\tablenotetext{d}{The model-derived radii are obtained using the measured age and mass. The $\rm T_{eff}$ and log g are calculated using individual radius measurements and measured mass.}
\end{table*}

\subsection{Measured Properties of \Targetname~B}

The Hydrogen-Burning Limit (HBL) is the minimum mass required for thermonuclear fusion in a star's core; it distinguishes brown dwarfs from low-mass stars. The accurate delineation of the HBL relies on factors like the equation of state, rotation, composition, and atmospheric attributes \citep{Burrows_1997}, generally falling within the range of 70-80 $\Mjup$. If adopting 75$\Mjup$ as the HBL, 14$\%$ of our dynamical mass posterior distribution falls below the HBL, and 86$\%$ lies above this value. If we instead adopt \citet{Chabrier_2023}'s new estimate for the HBL of 78.5$\Mjup$ that accounts for previously neglected physical effects, then 31$\%$ of our MCMC mass posterior lies below the substellar boundary. Alternatively, using the dynamical mass posterior from a narrow prior of $1.41 \pm 0.05\,M_\odot$ for the primary star, 11$\%$ and 38$\%$ of the mass posterior are below a HBL of 75 and 78.5 $\Mjup$, respectively. Our dynamical mass of ${81.9}_{-5.8}^{+6.4} \Mjup$ at 68$\%$ confidence cannot definitively classify it as either a brown dwarf or a low-mass star. However, the luminosity of \Targetname~B, inferred from its $L'$ photometry, suggests that the color of \Targetname~B more closely aligns with a substellar nature $-$ specifically, indicating a massive brown dwarf near the L/T transition.

The orbital eccentricities of substellar companions also shed light on key processes during their formation. Recent population-level inferences of 27 long-period giant planets and brown dwarfs (5–100 AU) by \citet{Bowler_2020} and \citet{Nagpal_2023} reveal distinct eccentricity distributions for giant planets (e$\approx$0.05–0.25) and brown dwarfs (e$\approx$ 0.6-0.9). This dichotomy suggests different formation channels: planets may form within protoplanetary disks, while brown dwarf genesis may occur through cloud fragmentation, resembling the formation channel of wide stellar binaries. \Targetname~B, orbiting at $\sim$20 AU with a tightly constrained eccentricity of $0.260_{-0.059}^{+0.065}$, exhibits a lower eccentricity than the broad peak toward high values (e$\approx$ 0.6-0.9) found for 18 imaged brown dwarf companions in \citet{Bowler_2020}. Further, they found that brown dwarf companions on closer orbits (5–30 AU) have a lower eccentricity distribution peak at e$\approx$0.5 compared to those at wider separations (30–100 AU), which peak at e$\approx$0.74. By both standards, \Targetname~B has a notably lower eccentricity. One possibility could be the presence of additional companions that circularized its orbit, as observed in stable, multi-planet systems like HR~8799 \citep{Bowler_2020}. In any scenarios, \Targetname~B provides a valuable data point for such eccentricity studies of directly imaged brown dwarfs.

Our orbital inclination also suggests that \Targetname~B's orbital orientation is close to being face-on as observed from the Earth. The near-face-on orbit would explain the significant astrometric acceleration of 64$\sigma$ in the HGCA, but subdued RV signal despite it being a massive candidate. We note that this is due to selection biases in the astrometric detection and recovery with imaging of companions found using accelerations. Continued direct imaging monitoring will enhance the precision of the companion's measured mass and orbit. 

\subsection{Comparison to Evolutionary Models}

\begin{figure*}
    \centering
\includegraphics[width=\textwidth]{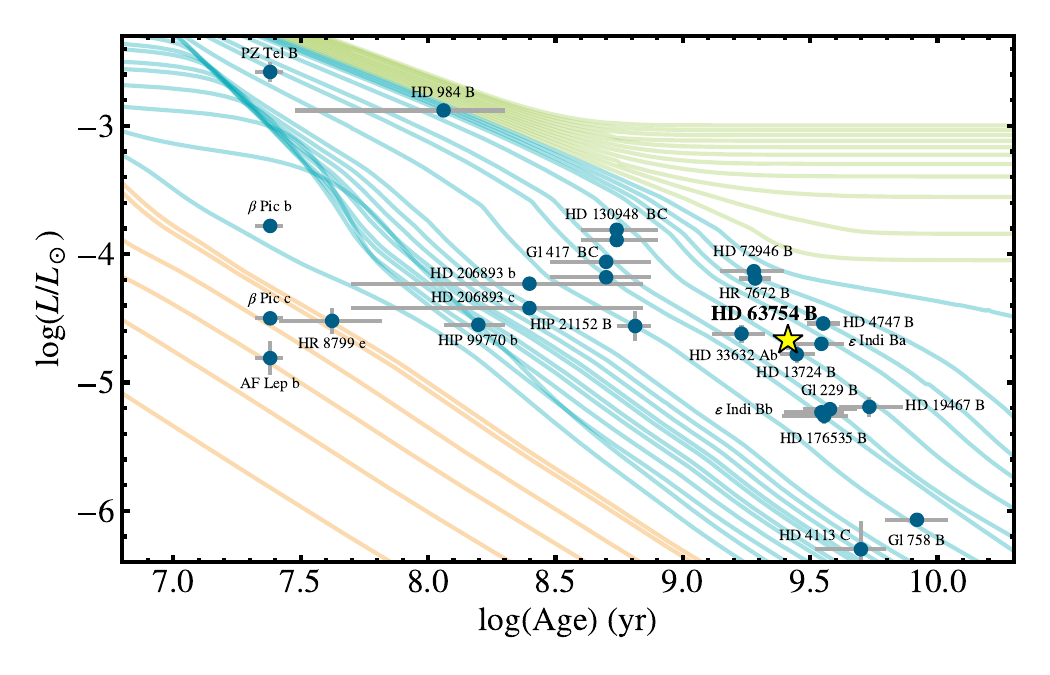}
    \caption{Imaged planets and brown dwarfs with dynamical masses, together with models of the luminosity evolution of planets (orange), brown dwarfs (blue), and low-mass stars (green) across various mass ranges \citet{Burrows_1997}. Our companion \Targetname~B is highlighted by the yellow star using the lower limit on the age of $>3.4$ Gyr and a dynamical mass of $81.9^{+6.4}_{-5.8} \Mjup$. The age uncertainty is not shown. Luminosity and age data for the majority of systems are tabulated in \citet{Franson_2023} with additional systems noted in \citep{Franson_2023b}. \label{all_bd_stats}}
\end{figure*}

We compare the independent mass, age, and luminosity measurements of \Targetname~B with three models of substellar evolution that provide self-consistent calculations for atmospheric structure and evolution across varying assumptions of atmospheric chemistry and boundary conditions. We consider the cloudless model TUSCON by \citet{Burrows_1997}, the cloudless ATMO-2020 model by \citet{Phillips_2020}, and the \citet{Saumon_2008} hybrid model with a mixture of cloud prescriptions (no clouds, hybrid, and cloudy). The \citet{Burrows_1997} TUSCON model uses lower-opacity `gray' atmospheres at higher temperatures, particularly intended for cloud-free cooler brown dwarfs below 1300 K. The \citet{Phillips_2020} ATMO-2020 model is an updated cloud-free model compared to TUSCON, following the same lineage as the ``Cond" \citep{Baraffe_2003} and ``BHAC15" \citep{Baraffe_2015} models. Both the TUSCON and ATMO 2020 cloudless models are designed specifically for cloud-free, later-type T dwarfs like Gl 229 B and Gl 758 B, and their effectiveness diminishes across the entire effective temperature range. On the other hand, the hybrid model does capture the transitional phase from L- to T-type cool brown dwarfs using the atmospheric model from \cite{Ackerman_2013} by varying the cloud sedimentation parameter. Particularly noteworthy is their discovery that the transition from cloudy L dwarfs to cloudless T dwarfs results in a deceleration of the evolutionary process, leading to an accumulation of substellar objects in the L/T transition phase. This finding is corroborated by several other recent studies such as \citet{Chen_2022} who found evidence of slowed cooling on the L/T transition. 

We use the method outlined in \citet{Dupuy_2019} to infer masses of \Targetname~B based on its measured luminosity and age obtained from each evolutionary model. We adopt the broad age posterior in Figure~\ref{age_posterior} with a lower limit of 3.4 Gyr, and a luminosity of $\log(L_{\rm bol}/\Lsun)=-4.55\pm0.08\,$dex as previously derived. Briefly, we use an importance sampling approach to calculate inferred masses for each evolutionary model drawing random samples from the distributions of bolometric luminosity and age. Then, we bilinearly interpolate the model grid to determine the companion's mass corresponding to the given age and luminosity values. We repeat the process to build a model-derived mass distribution from each model. The inferred mass distributions for the three evolutionary grids are depicted in Figure~\ref{mass_comparison}, along with our measured dynamical mass distribution of \Targetname~B from the orbital fit. In a similar fashion, we use this importance sampling technique to estimate the posterior distributions for age with known MCMC mass posterior and our inferred luminosity, and for luminosity with known masses and age. 

We quantify the differences between the model-independent measurements of mass, age and luminosity ($\rm X_{model-inferred}$) and model-inferred masses ($\rm X_{empirical}$) by computing the probability P($\mathrm{X_{model-inferred} > X_{empirical}}$). The discrepancy in model-derived and model-independent values can be converted to a one-sided Gaussian-equivalent
standard deviation via the special error function:
\begin{equation}
\mathrm{\sigma = \sqrt{2} erf^{-1}(1 - 2 P(X_{model-inferred} > X_{empirical})})
\end{equation}
We list the inferred parameters including mass, age, luminosity, effective temperature, radius and surface gravity extracted from different evolutionary models in Table~\ref{model_derived_params}. 

Here, we compare the
model-inferred parameters with empirically measured mass, age and
luminosity of HD 63754 B. Figure~\ref{mass_comparison} shows the model-independent and model-inferred histograms for mass, age and luminosity as sourced from the TUCSON, the ATMO 2020 and the SM08 hybrid models, respectively. The measured and inferred parameters are consistent within 3$\sigma$ except for the agreement between the hybrid model masses and the dynamical mass (4$\sigma$). Figure~\ref{evol_grid} illustrates the position of \Targetname~B with respect to the predicted masses and ages from each of the three evolutionary models. The \citet{Burrows_1997} TUCSON model emerges as the closest match to our measured parameters for \Targetname~B, providing an inferred mass of $75.0_{-3.0}^{+5.3} \Mjup$ about 1.5$\sigma$ away from our dynamical mass measurement given the measured age and luminosity. However, we note that this observed better agreement with the TUCSON grid is likely a coincidental outcome rather than indicative of a better fit. Several important caveats must be considered when interpreting these comparisons. First, the TUCSON evolutionary model grids are specifically calibrated for effective temperatures ranging from 125 to 1200 K, which are applicable only to T dwarfs and colder brown dwarfs. In contrast, the newer ATMO 2020 and SM08 hybrid models are more appropriate for L/T transition brown dwarfs like \Targetname~B, covering a broader effective temperature range of 700 to 2000 K. Furthermore, the cloudless TUCSON model being the best-fit model contradicts \Targetname~B's L/T transition spectral type determination. The slightly above-solar metallicity observed in \Targetname~B's host star also argues against a scenario of low opacity and a cloudless model, suggesting instead the presence of a cloudy atmosphere that could influence its thermal evolution.

Our measured and model-inferred parameters presented in Table~\ref{model_derived_params}, and by extension the comparison between them, could also be biased by additional sources of error. The spectral typing method employed here, which relies on color relations, is less accurate than empirical spectral fitting. This is exemplified by our recent work on HD~176535~B (HIP~93398~B) \citep{Li_2023}, which was mistakenly classified as a T6 brown dwarf using single-band photometry but was later revised to a late-L brown dwarf based on follow-up spectra (Lewis et al. 2024, submitted). Additionally, the dynamical mass of \Targetname~B exceeds the mass range of substellar models considered here (typically $\le 0.075 \Msun \approx 78.6 \Mjup$), potentially introducing systematic uncertainties in the model-inferred parameters. 

Figure~\ref{all_bd_stats} shows a comprehensive list of imaged brown dwarfs with independently measured age, mass and luminosity. The location of \Targetname~B is shown by the yellow star. While the majority of benchmark brown dwarfs with independent simultaneous age, mass, and luminosity measurements generally agree with evolutionary model predictions (e.g. HD~4747~B \citep{Peretti_2019}, HD~33632~Ab \citep{Currie_2021}, Gl~758~B \citep{Bowler+Dupuy+Endl+etal_2018}), a handful of brown dwarf systems bear pronounced discrepancies with evolutionary models. Previous statistical investigations \citep{MBrandt_2021, Dupuy_2019,Li_2023,Franson_2023} on benchmark brown dwarfs underscore two emerging trends: 1) some measured brown dwarfs are over-luminous in their early stages of evolution 2) others are under-luminous at high masses and old ages, in tension with their observed spectral types. 

The first trend, or the under-luminosity problem at young substellar ages, exemplified by HD~130948~BC \citep{Dupuy_2009}, has been poorly explained. Most evolutionary models suggest that brown dwarfs could not have cooled to the observed luminosity at young ages. Similarly, their measured masses are unexpectedly low, or the ages of their host stars are older than the cooling ages of substellar objects. Such under-massive systems include HD~13724~B (4$\sigma$ age discrepancy) \citet{Rickman_2020}, HD~130948~BC (3$\sigma$), Gl~417~BC (1$\sigma$) \citep{Dupuy_2014}, and CWW~89~Ab (7$\sigma$). While a variety of mechanisms have been considered, including inhomogeneous dust clouds, non-equilibrium chemistry, thermal inversion, magnetic activity, non-solar metallicity, vertical mixing, and cloud formation, none have provided a comprehensive account of this phenomenon, although thermal inversion could partially explain the slowed cooling of specific L and T brown dwarfs like CWW~89~A \citep{Bardalez_Gagliuffi_2014}. Atmospheric retrieval of L-type brown dwarfs has revealed upper atmospheric heating or similar surface processes \citep{Burningham_2017} may contribute partially to the over-luminosity problem, but the major causes are still unclear to date.

The latter trend, known as the over-luminosity brown dwarf problem, has several possible explanations. This problem is now evident in a significant number of brown dwarfs, including Gl~229~B \citep{Brandt_2020}, HD~19467~B \citep{Maire_2020}, HD~47127~B \citep{Bowler_2021}, HD~176535~B \citep{Li_2023}, and HD 4113~C \citep{Cheetham_2018}. The first possible resolution for this discrepancy can be attributed to unresolved multiplicity in these brown dwarf systems. Multiplicity studies find a relatively high binary fraction of 8$\pm 6\%$ for ultracool T8-Y0 brown dwarfs \citep{Fontanive_2018}. The discrepancy may also arise from inherent inaccuracies in our orbit modeling using Hipparcos-Gaia proper motion anomaly, especially in the high sigma regime. More accurate solutions are expected with the upcoming Gaia DR4 release in 2025. We discuss this in more detail in section~\ref{gaiadr4_accl}. If the over-massive issue cannot be ascribed to unresolved multiplicity or inaccuracies in astrometric data, it may indicate a deeper problem with the evolutionary models and their underlying assumptions. Rectifying this situation would likely require substantial overhauls in how these models handle fundamental properties such as the equation of state, thermal transport in the interior structure, or the effects of rotation and magnetism \citep{Marley_2021}.

One of the earliest discovered and coolest brown dwarfs, Gl~229~B, is a significant outlier in this over-luminosity regime. Its mass and luminosity significantly differ from evolutionary model predictions. Neither low metallicity nor the presence of a massive, inner companion can address this discrepancy. The likely explanation based on multi-decade observations could be that Gl~229~B itself is an unresolved binary system \citep{MBrandt_2021} with a total system mass of 71.4 $\Mjup$. Although the departure from evolutionary model predictions is not as severe as are the cases with Gl~229~B (10$\sigma$) and HD 4113~C (5$\sigma$), \Targetname~B still falls into this category with a discrepancy at levels as large as 3$\sigma$, even at older ages. 
We list several possibilities for the extra mass in the \Targetname~AB system. 1) \Targetname~B has a low-mass-ratio binary companion, whose flux would contribute negligibly to the integrated spectrum or the infrared photometry of \Targetname~B. 2) There are additional companions in the \Targetname~AB system. We find no evidence of any wide, massive companions in our Keck/NIRC2 AO imaging as shown in Figure~\ref{Annular_Psf}, but we cannot make the conclusion with a single night data over only about an hour integration time. 
Our orbital fit to the RV and absolute astrometric data in Figure~\ref{fig:orbit_fit_plots} also cannot rule out disturbance to the system from a less massive, inner companion or a binary companion to \Targetname~B. The relative orbit has a high scatter and relies on only a single astrometry point which leaves some space for either an unresolved binary companion around \Targetname~B or additional exoplanets in the \Targetname~AB system. Further high-resolution imaging with smaller inner working angles and spectroscopic observations may identify any potential additional companions. 3) \Targetname~B is not an isolated incidence but rather an indication of a systematic problem with evolutionary models \citep{MBrandt_2021}. 

\subsection{Predicted Accelerations for Gaia DR4}
\label{gaiadr4_accl}

\begin{figure}
    \centering
    \includegraphics[width=0.5\textwidth]{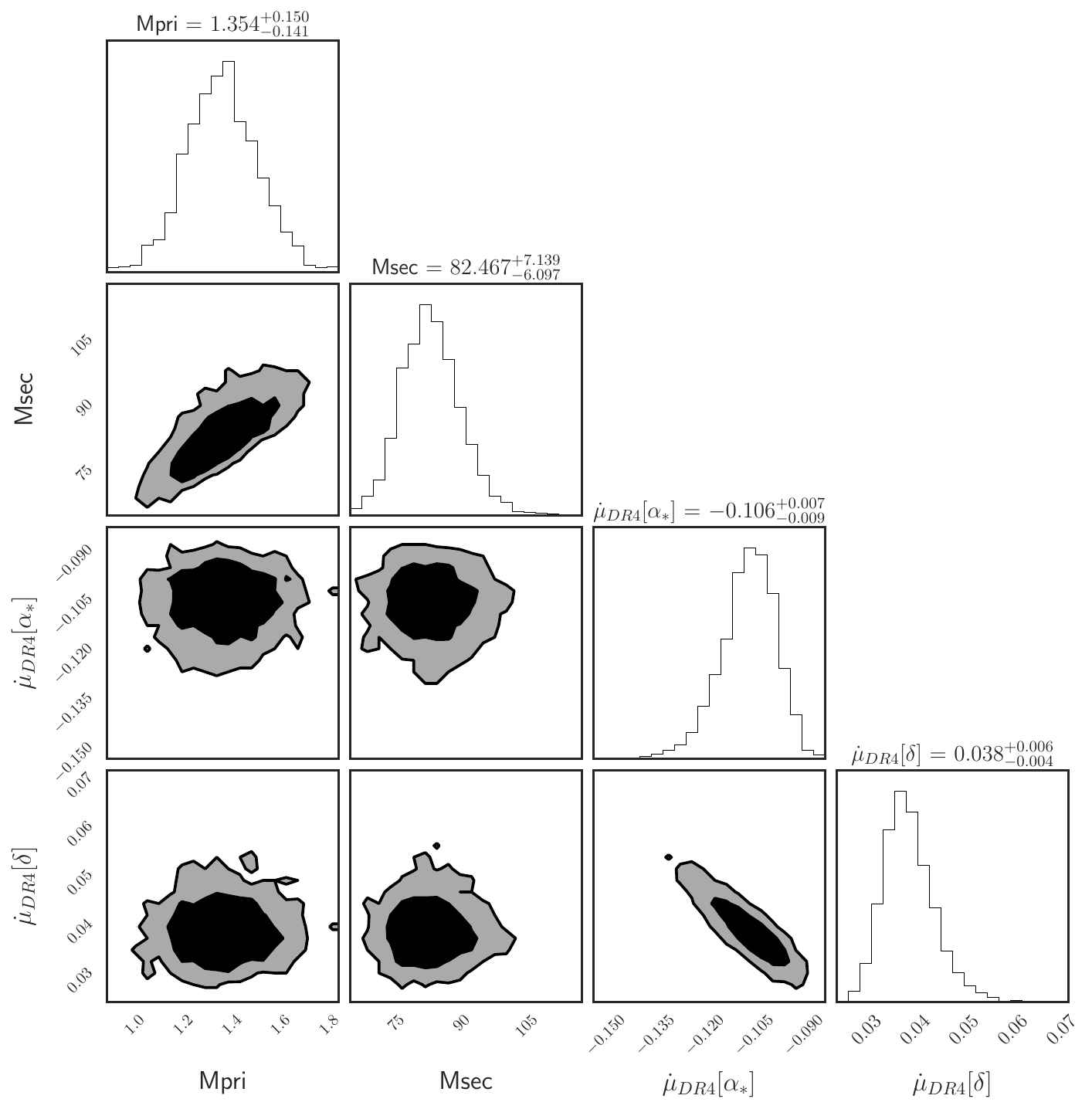}
    \caption{Corner plot showing the mass distribution of the \Targetname~AB system and the projected accelerations from the orbital fit at the Gaia DR4 central epoch of 2017.5. No strong correlations are seen between the companion mass and DR4 predicted accelerations based on current measurements, which means that DR4 may not be as useful in further constraining the dynamical mass of \Targetname~B. \label{DR4_predictions}}
\end{figure}

Gaia Data Release 4 (DR4), scheduled for release in 2025, will offer high-precision individual epoch astrometry and a 5-year time baseline, providing more precise constraints on planetary orbits. We forward-model \Targetname~A's predicted stellar accelerations at Gaia DR4's central epoch of 2017.5 using projections from our MCMC \orvara result for \Targetname~B. We employ an adaptation of the epoch astrometry fitting code $\tt htof$ \citep{Brandt_2021c}, which is an open-source tool to fit arbitrary high-order astrometric solutions to epoch astrometry. The predicted astrometric acceleration caused by \Targetname~B at 2017.5 is $\dot{\mu}_{DR4, \alpha_{*}} = -0.106_{-0.009}^{+0.007}$ m/s/yr and $\dot{\mu}_{DR4, \delta} = 0038_{-0.004}^{+0.006}$ m/s/yr, with the uncertainty coming from the range of orbits compatible with existing data. Figure~\ref{DR4_predictions} shows a corner plot of the masses of the two known bodies in the \Targetname system and their correlation with the predicted Gaia DR4 accelerations. The correlation coefficients between DR4 accelerations and the mass of \Targetname~B are close to zero, suggesting that Gaia DR4 may not be as useful in refining \Targetname~B's dynamical mass and validating our orbital solution. This is unsurprising given the large separation of this massive companion. Gaia DR4 will be sensitive to short period companions which will be useful if they can also be found with speckle imaging. For now, whether there are additional companions in the \Targetname~AB system remains unresolved. Further RV, direct imaging and spectroscopic monitoring will clarify the nature of this system.

\section{Conclusion}
\label{sec:conclusion}
In this study, we presented the joint astrometric and direct imaging discovery of \Targetname~B, a companion selected for observation in our Keck-HGCA accelerating stars program because of its significant \Hipparcos-\Gaia astrometric signature. We carried out comprehensive orbital and evolution modeling using empirical measurements of stellar velocities, activity-age indicators, direct imaging photometry, and evolutionary theoretical isochrones. The derived dynamical mass for \Targetname~B is $81.9_{-5.8}^{+6.4} \Mjup$, with an inferred activity age $>$3.4 Gyr and a bolometric luminosity of $\mathrm{log(L_{bol}/\Lsun)= -4.55 \pm0.08}$ dex. Comparison of our findings regarding dynamical mass, age, and luminosity with substellar evolutionary models reveals that \Targetname~B exhibits characteristics of being over-massive and under-luminous relative to model-predicted outcomes. We conclude that such discrepancies could stem from uncertainties in stellar age determination, potential systematic biases in evolutionary models, or unresolved multiplicity within the system.

\Targetname~B, alongside several other objects, poses a challenge to the current understanding of substellar cooling models. While the precise nature of \Targetname~B remains unclear, its L' band photometry provides support for a substellar classification, particularly favoring a brown dwarf on the L/T transition. We anticipate that additional orbital or spectroscopic measurements will finally clarify the nature of \Targetname~B. Enhanced observational data for a broader sample of brown dwarfs discovered through acceleration could significantly refine our comprehension of the substellar evolution paradigm.

\section{Acknowledgement}

This research utilized data from the Keck Observatory Archive (KOA), operated jointly by the W. M. Keck Observatory and the NASA Exoplanet Science Institute (NExScI), under contract with the National Aeronautics and Space Administration (NASA). The findings presented herein made use of data from the European Space Agency (ESA) Gaia space mission, with data processing conducted by the Gaia Data Processing and Analysis Consortium (DPAC). Financial support for the DPAC is provided by national institutions, particularly those institutions participating in the Gaia MultiLateral Agreement (MLA). The Gaia mission website is \url{https://www.cosmos.esa.int/gaia} and the Gaia archive can be accessed at \url{https://archives.esac.esa.int/gaia.} T.D.B.~gratefully acknowledges support from the Alfred P.~Sloan Foundation and from the NASA Exoplanet Research Program under grant \#80NSSC18K0439. B.L.L. acknowledges support from the National Science Foundation Graduate Research Fellowship under Grant No. 2021-25 DGE-2034835. Any opinions, findings, and conclusions or recommendations expressed in this material are those of the authors(s) and do not necessarily reflect the views of the National Science Foundation. B.P.B. acknowledges support from the NASA Exoplanet Research Program grant 20- XRP20$\_$2-0119 and the Alfred P. Sloan Foundation. R.K. acknowledges the support by the
National Science Foundation under Grant No. NSF PHY-1748958. We thank the Heising Simons Foundation for their support.

\section*{Data Availability}
This manuscript includes data acquired by the NIRC2 camera at the W. M. Keck Observatory (WMKO), accessible to the public via the Keck Observatory Archive (KOA), which is jointly managed by the WMKO and NASA Exoplanet Science Institute (NExScI). This research is funded by the National Aeronautics and Space Administration. All data in this paper are publicly available in the KOA. The data are analyzed with the $\orvara$ and $\VIP$ open-source packages, which are publicly available at \url{https://github.com/t-brandt/orvara} and \url{https://github.com/vortex-exoplanet/VIP}, respectively. We also acknowledge the use of public Gaia EDR3 data data through the Gaia Archive at \url{https://gea.esac.esa.int/archive/}.

\bibliographystyle{apj_eprint}
\bibliography{refs}

\label{lastpage}

\end{document}